\documentclass[twocolumn,english,aps,superscriptaddress,showpacs,prl,floatfix,nofootinbib,longbibliography]{revtex4-1}
\usepackage{mathptmx}

\usepackage[latin9]{inputenc}
\usepackage{color}
\usepackage{babel}
\usepackage{amsmath}
\usepackage{amssymb}
\usepackage{graphicx}
\usepackage{esint}
\usepackage[unicode=true,pdfusetitle,
 bookmarks=true,bookmarksnumbered=false,bookmarksopen=false,
 breaklinks=true,pdfborder={0 0 0},pdfborderstyle={},backref=false,colorlinks=true]
 {hyperref}
\hypersetup{
 linkcolor=blue,citecolor=blue,urlcolor=blue}
\usepackage{breakurl}

\makeatletter
\usepackage{dsfont}

\makeatother

\begin{document}

\title{Disorder-induced dynamical Griffiths singularities after certain
quantum quenches}

\author{Jos\'e A. Hoyos\,\href{https://orcid.org/0000-0003-2752-2194}{\includegraphics[keepaspectratio,width=0.7em]{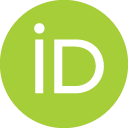}}}

\affiliation{Instituto de F\'{\i}sica de S\~ao Carlos, Universidade de S\~ao
Paulo, C.~P.~369, S\~ao Carlos, S\~ao Paulo 13560-970, Brazil}

\affiliation{Max Planck Institute for the Physics of Complex Systems, N\"othnitzer
Str. 38, 01187 Dresden, Germany}

\author{R. F. P. Costa\,\href{https://orcid.org/0000-0001-9815-8727}{\includegraphics[keepaspectratio,width=0.7em]{orcid.png}}}

\affiliation{Instituto de F\'{\i}sica, Universidade Federal de Uberl\^andia,
C.~P.~593, 38400-902 Uberl\^andia, MG, Brazil}

\author{J. C. Xavier\,\href{https://orcid.org/0000-0003-4913-3480}{\includegraphics[keepaspectratio,width=0.7em]{orcid.png}}}

\affiliation{Instituto de F\'{\i}sica, Universidade Federal de Uberl\^andia,
C.~P.~593, 38400-902 Uberl\^andia, MG, Brazil}
\begin{abstract}
We demonstrate that in a class of disordered quantum systems the dynamical
partition function is not an analytical function in a time window
after certain quantum quenches. We related this behavior to rare and
large regions with atypical inhomogeneity configurations. We also
quantify the strength of the associated singularities and their signatures
in experiments and numerical studies.\\
\\
Published in \href{https://doi.org/10.1103/PhysRevB.106.L140201}{Phys. Rev. B {\bf 106}, L140201 (2022)};
DOI: \href{https://doi.org/10.1103/PhysRevB.106.L140201}{10.1103/PhysRevB.106.L140201}
\end{abstract}

\date{\today}

\maketitle
Phase transitions (PTs) are among the most intriguing phenomena in
nature. When crossed, the macroscopic properties of matter change
fundamentally, often requiring new concepts for a proper description~\citep{fisher-book-1965}.
In thermodynamic equilibrium, PTs are on firm theoretical grounds
as they occur whenever a zero of the partition function touches the
real-temperature (or field) axis~\citep{yang-lee-pr52,fisher-book-1965}.
Consequently, thermodynamic observables become non-analytic functions
of temperature/field at the transition point. Notably, these Yang-Lee-Fisher
(YLF) zeros were recently measured experimentally~\citep{peng-etal-prl15,brandner-etal-prl17}.

Inhomogeneities, which are nearly ubiquitous in experiments, play
an important role in equilibrium PTs. For instance, even the smallest
amount of them can change the singularities of a critical system~\citep{harris-jpc74},
smear the PT~\citep{vojta-prl03,hoyos-vojta-prl08}, or even destroy
it~\citep{imry-ma-prl75}. Another remarkable inhomogeneity-induced
phenomenon is the stabilization of a Griffiths phase (GP): an extended
region in the phase diagram surrounding a phase-transition manifold
where the free energy is non-analytic~\citep{griffiths-prl69,mccoy-prl69}.
Counterintuitively, the non-analyticity is due to so-called rare regions
(RRs)\textemdash large and rare regions in space with atypical configurations
of inhomogeneities\textemdash which provide YLF zeros arbitrarily
close to the real-temperature axis~\citep{griffiths-prl69,wortis-prb74,harris-prb75}. 

Over the past decades, the influence of the RRs on many observables
has been quantified in a multitude of strongly interacting systems
ranging from classical and quantum models in equilibrium to non-equilibrium
reaction-diffusion models (for reviews, see Refs.~\citealp{igloi-review,vojta-review06,igloi-monthus-review2}).
In the associated GPs, the RRs endow many observables with singular
behavior in the long-time/low-frequency regime. This common feature
is due to the RRs' long relaxation times~\citep{randeria-etal-prl85,bray-prl87,bray-prl88,thill-huse-physa95,vojta-hoyos-prl14}. 

With the growing capacity of experimentally accessing the time evolution
of closed quantum many-body systems~\citep{bloch-dalibard-zwerger-rmp08,georgescu-etal-rmp14},
it then became natural to inquire whether the RRs play any important
role in their time evolution. Clearly, the notion of slow RRs at equilibrium
does not apply and thus their importance cannot be anticipated. Evidently,
obtaining a result on the RR effects in a general out-of-equilibrium
situation is desirable but very unlikely to exist. We thus restrict
ourselves to the simpler case of quantum quenches which already allows
the study of fundamental phenomena such as entanglement spreading
and thermalization~\citep{calabrese-cardy-prl06,polkovnikov-etal-rmp11,mitra-arcmp18}.
Here, the system's initial state is $\left|\psi_{0}\right\rangle $,
the ground state of $H_{0}\equiv H\left(h_{0}\right)$, and time evolved
according to the postquench Hamiltonian $H\equiv H\left(h\right)$,
with $h$ being a tuning parameter. In this context, the concept of
dynamical quantum phase transitions (QPTs) is quite useful~\citep{heyl-etal-prl13}
because an analogy with equilibrium PTs can be made. The linking quantity
is the dynamical free energy 
\begin{equation}
f(t)=-V^{-1}\ln\left|Z(t)\right|^{2},\mbox{ where }Z(z)=\left\langle \psi_{0}\left|e^{-iHz}\right|\psi_{0}\right\rangle \label{eq:Z}
\end{equation}
 is the return probability amplitude after the quench, $z=t+i\tau$
is the complex time, and $V$ is the system volume. $Z$ is the dynamical
analog of the equilibrium partition function. As in equilibrium PTs,
its zeros accumulate in lines or areas on the complex-time plane and,
in the thermodynamic limit, may touch the real-time axis. When this
happens, a dynamical QPT occurs~\citep{heyl-etal-prl13,andarschko-sirker-prb14,vajna-dora-prb14,schmitt-kehrein-prb15,halimeh-stauber-prb17,zunkovic-etal-prl18,jafari-sr-19}
and has been experimentally verified in different quantum simulator
platforms~\citep{jucervic-etal-prl17,zhang-etal-nature17,bernie-etal-nature17,flaschner-etal-np18,guo-etal-pra19}
(for a review, see Ref.~\citealp{heyl-rpp18}).

\begin{figure}
\begin{centering}
\includegraphics[clip,width=0.9\columnwidth]{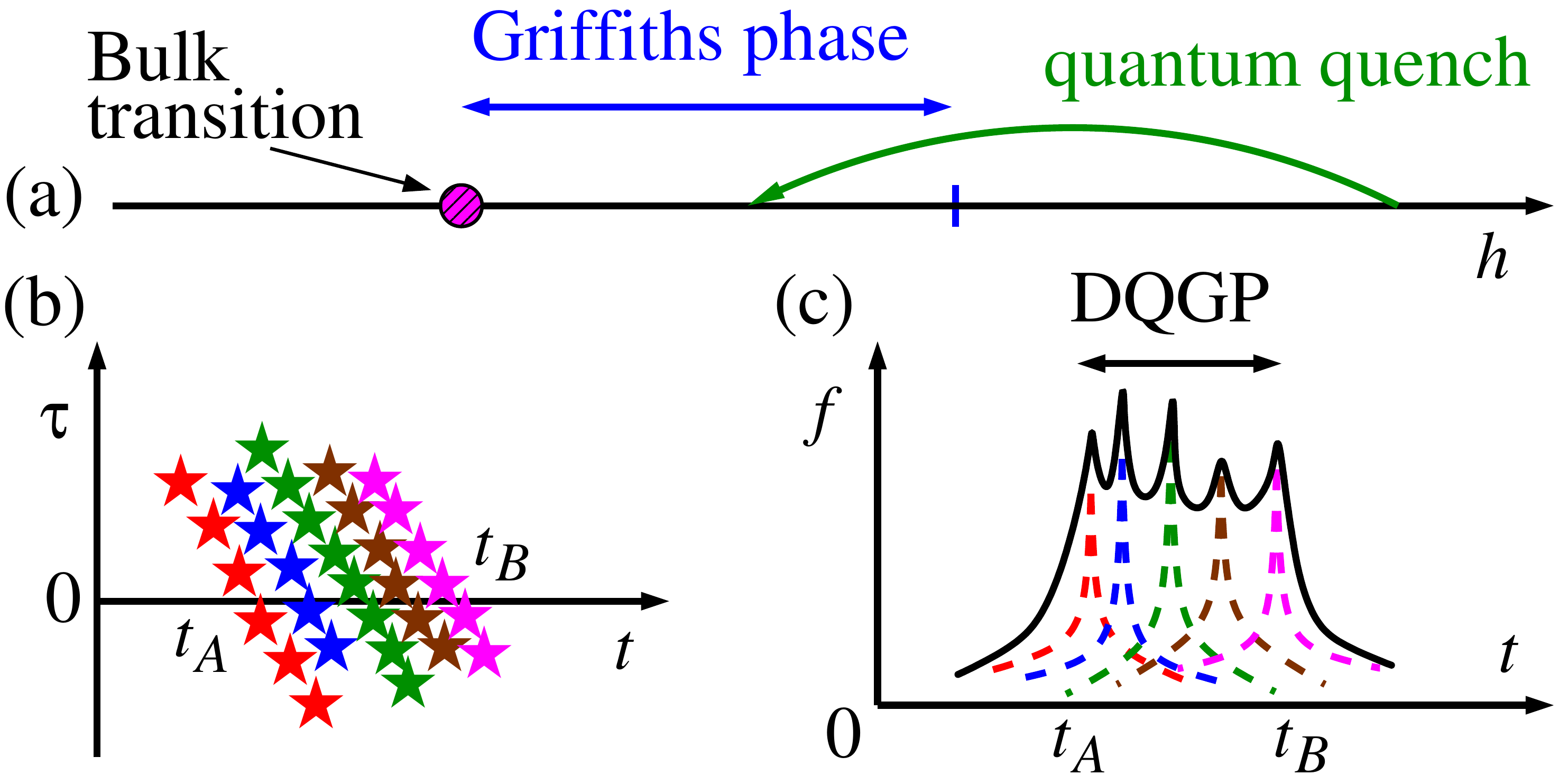}
\par\end{centering}
\caption{Schematics of (a) the equilibrium phase diagram and the class of quantum
quenches studied: from a point far from the quantum phase transition
to a point in the nearby Griffiths phase. (b) The rare-region-induced
zeros of the dynamical partition function $Z$, and (c) the associated
dynamical free energy $f$ (solid line). Each set of zeros (stars
of a given color) and the corresponding dynamical free energy $f_{\text{RR}}$
(dashed line) are due to a single rare region. The singular part of
$f$ is a simple superposition of all $f_{\text{RR}}$. The time window
$t_{A}<t<t_{B}$ is the dynamical quantum Griffiths phase.\label{fig:summary}}
\end{figure}

In this Letter, we use the unifying concept of YLF zeros to show that
the RRs dominate the system's early-time dynamics for all quenches
which do not cross the bulk equilibrium QPT but do cross the RR local
QPT, i.e., the quantum quenches are from a conventional phase to the
nearby GP {[}see Fig.~\hyperref[fig:summary]{\ref{fig:summary}(a)}{]}.
For those quenches, the RRs endow $Z(z)$ with YLF zeros arbitrarily
close to the real-time axis. As in equilibrium GPs, these YLF zeros
are spread over an area on the complex-time plane with the associated
density of zeros depending on the details of the disorder variables
in $H$ {[}see Fig.~\hyperref[fig:summary]{\ref{fig:summary}(b)}{]}.
We thus propose the term dynamical quantum Griffiths phase to designate
the real-time axis interval intersected by the YLF zeros {[}see Fig.~\hyperref[fig:summary]{\ref{fig:summary}(c)}{]}.

The reasoning behind our result is as follows. After the quench, the
bulk remains nearly in its ground state since its QPT was not crossed.
The RRs, however, are highly excited. Because the RRs and the bulk
are in different phases, these excitations do not rapidly decay. Thus,
meanwhile, the RRs' dynamics is decoupled from the bulk's in a sense
that will become precise later. Consequently, two sets of YLF zeros
appear, one provided by the bulk and the other by the RRs. Those from
the bulk are far from the real-time axis and thus only provide analytical
contributions to $f(t)$. Those from the RRs, however, are arbitrarily
close to the real-time axis and therefore are responsible for the
non-analyticities of $f(t)$. In addition, we show that this singular
behavior can be well approximated by that of completely decoupled
RRs with open boundary conditions undergoing the same quantum quench.

We remark that, differently from the known cases in the literature,
the RRs in dynamical QPTs dominate the short-time dynamics. This is
exciting because it allows for an easier identification of the RRs'
effects in numerical studies and in experiments. 

Finally, we notice that quenched disorder effects on dynamical QPTs
were studied in a variety of models~\citep{obuchi-takahashi-pre12,yang-etal-prb17,yin-etal-pra18,gurarie-pra19,cao-etal-prb20,mishra-jafari-akbari-jpa20}.
These studies, however, did not focus on the RR-induced effects.

In the remainder of this Letter, we derive our results from an explicit
model Hamiltonian, discuss their generality and extensions, and provide
concluding remarks.

Consider the transverse-field Ising chain 
\begin{equation}
H=-\sum_{i=1}^{L}J_{i}\sigma_{i}^{z}\sigma_{i+1}^{z}-h\sum_{i=1}^{L}\sigma_{i}^{x},\label{eq:H}
\end{equation}
 where $\boldsymbol{\sigma}_{i}$ are Pauli matrices, $J_{i}>0$ are
the ferromagnetic coupling constants (which, due to inhomogeneities,
are site dependent), and $h>0$ is the transverse field and plays
the role of the tuning parameter of $H\left(h\right)$. We consider
chains of $L$ sites long with periodic boundary conditions $\boldsymbol{\sigma}_{L+i}=\boldsymbol{\sigma}_{i}$.
The model has two zero-temperature phases: the ferromagnet ($h<h_{c}$)
and the paramagnet ($h>h_{c}$) separated by a quantum critical point
at $h_{c}=J_{\text{typ}}$, where $J_{\text{typ}}=e^{\overline{\ln J}}$
is the geometric mean of the coupling constants~\citep{pfeuty-pla79}.

The clean system ($J_{i}=J$) can be solved analytically using standard
methods~\citep{supp-mat}. The return probability amplitude \eqref{eq:Z}
after the quantum quench $h_{0}\rightarrow h$ (with $h_{0}>h_{c})$
is 
\begin{equation}
Z(z)=e^{-iE_{0}z}\prod_{0<k_{n}<\pi}\left(1-\frac{\left(1-\varepsilon_{k_{n}}\right)\left(1-e^{-4i\omega_{k_{n}}\left(h\right)z}\right)}{2}\right),\label{eq:Z-clean}
\end{equation}
 where $E_{0}=-\sum_{n=1}^{L}\omega_{k_{n}}\left(h\right)$ is the
ground-state energy of the post-quench Hamiltonian, the momenta $k_{n}=\left(2n-1\right)\frac{\pi}{L},$
$n=1,\dots,L,$ $\omega_{k}\left(h\right)=\sqrt{h^{2}-2hJ\cos k+J^{2}}$
is the dispersion relation, and $\varepsilon_{k}=\varepsilon_{k}(h,h_{0})\equiv\left[hh_{0}-J\left(h_{0}+h\right)\cos k+J^{2}\right]/\left[\omega_{k}\left(h\right)\omega_{k}\left(h_{0}\right)\right]$.
The YLF zeros of \eqref{eq:Z-clean}, $z^{*}=t^{*}+i\tau^{*}$, are 

\begin{equation}
t_{m,n}^{*}=\frac{\left(2m+1\right)\pi}{4\omega_{k_{n}}\left(h\right)}\text{ and }\tau_{n}^{*}=\frac{\ln\left(\frac{1+\varepsilon_{k_{n}}(h,h_{0})}{1-\varepsilon_{k_{n}}(h,h_{0})}\right)}{4\omega_{k_{n}}\left(h\right)},\label{eq:YLF-clean}
\end{equation}
 where $m\in\mathbb{N}$ defines different accumulation lines of zeros
(for a graphical illustration, see~\citep{supp-mat}). These lines
pierce the real-time axis if and only if the equilibrium QPT is crossed
by the quantum quench, i.e., iff $\left(h-h_{c}\right)\left(h_{0}-h_{c}\right)<0$
in the model \eqref{eq:H}. In the following, we numerically demonstrate
that even a single RR dramatically change this scenario. 

Unfortunately, there is no analytical solution for the non-homogeneous
case. We then compute $Z(z)$ in \eqref{eq:Z} via exact numerical
diagonalization and find its YLF zeros $z^{*}$ using the standard
secant method~\citep{supp-mat}. For definiteness, we set the couplings
in the Hamiltonian \eqref{eq:H} to $J_{i}=J_{\text{B}}$ (the bulk
couplings) everywhere except inside a RR where $J_{i}=J_{\text{RR}}$
for $1\leq i\leq L_{\text{RR}}-1$. The fact that we are considering
a compact RR is of no consequence for our purposes. Later, we discuss
more general profiles. For simplicity, we consider quantum quenches
from $h_{0}=\infty$ to a finite $h$. Thus, $\left|\psi_{0}\right\rangle =\otimes_{i=1}^{L}\left|\rightarrow\right\rangle $,
with $\sigma^{x}\left|\rightarrow\right\rangle =\left|\rightarrow\right\rangle $,
is a simple product state. We want to study quenches that do not cross
the bulk QPT, and thus $h>J_{\text{B}}$. In the following numerical
study, we set $h=5J_{\text{B}}$. Other values only produce quantitative
changes and will be shown elsewhere.

\begin{figure}[t]
\begin{centering}
\includegraphics[clip,width=1\columnwidth]{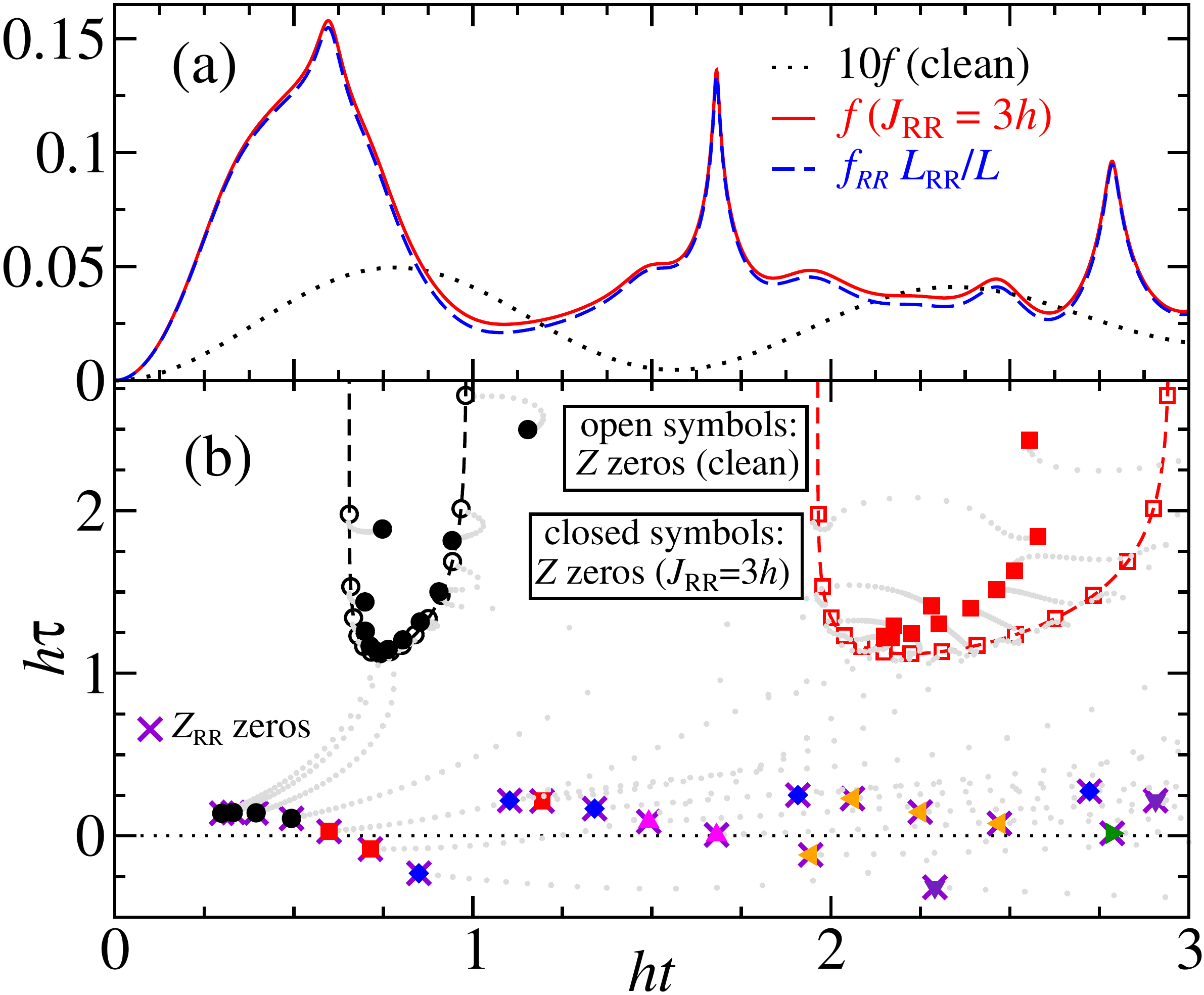}
\par\end{centering}
\caption{(a) The dynamical free energy $f$ as a function of the real time
$t$ for three different chains after the quantum quench from $h_{0}=\infty$
to finite $h$. The first chain (black dotted line) is homogeneous,
$L=30$ sites long with periodic boundary conditions, and has couplings
$J_{\text{B}}=h/5$. The second chain (red solid line) is identical
to the first one except that it contains a RR of size $L_{\text{RR}}=8$
inside which the couplings are $J_{\text{RR}}=3h$. The third chain
(blue dashed line) is homogeneous, $L_{\text{RR}}$ sites long with
open boundary conditions, and has couplings $J_{\text{RR}}$. (b)
The corresponding Yang-Lee-Fisher zeros of $Z(z)$ for these three
chains: open symbols, solid symbols, and $\times$ symbols, respectively.
The zeros' trajectories of the second chain (when changing $J_{\text{RR}}$
from $h/5$ to $3h$) are given by the gray dots (see text).\label{fig:YLF}}
\end{figure}

We show in Fig.~\hyperref[fig:YLF]{\ref{fig:YLF}(a)} the dynamical
free energy for the homogeneous case $J_{\text{RR}}=J_{\text{B}}$
for a chain of only $L=30$ sites long (for the sake of clarity) with
periodic boundary conditions. The resulting curve (dotted line; notice
it is multiplied by a factor of $10$) is completely smooth and analytic
as expected. The corresponding YLF zeros Eq.~\eqref{eq:YLF-clean}
are shown in Fig.~\hyperref[fig:YLF]{\ref{fig:YLF}(b)} as open symbols.
As is well known~\citep{heyl-etal-prl13}, they accumulate in lines
far from the real-time axis. For the time window considered, only
the first two accumulation (dashed) lines appear. Increasing $J_{\text{RR}}$
gradually (in steps of $0.1h$ up to $3h$ and considering, for the
sake of clarity, a rare region of only $L_{\text{RR}}=8$ sites long),
the zeros move on the complex-time plane {[}see gray dots in Fig.~\hyperref[fig:YLF]{\ref{fig:YLF}(b)}{]}.
Analyzing their trajectories, we verify two distinct sets of zeros:
one that remains in the upper half of the complex-time plane and the
other which migrates to the vicinity of the real-time axis. The latter
set of zeros accumulate in lines which pierce the real-time axis for
$J_{\text{RR}}>h$. For the case $J_{\text{RR}}=3h$, we plot the
corresponding $f(t)$ in Fig.~\hyperref[fig:YLF]{\ref{fig:YLF}(a)}
(red solid line). The corresponding zeros are shown in Fig.~\hyperref[fig:YLF]{\ref{fig:YLF}(b)}
as solid symbols. The developing singularities in $f(t)$ are in one-to-one
correspondence with the zeros close to the real-time axis. 

Our interpretation of the latter set of zeros is that the unitary
dynamics of the RR is essentially decoupled from the bulk. The reasoning
is as follows. The bulk is gapful and is locally in a different phase
from the RR. The RR excitations (kinks) have a different nature from
the bulk's (spin flips). Therefore, the quench-induced excitations
of the RR do not immediately decay into the bulk.

To give support to this interpretation, we compute the dynamical free
energy $f_{\text{RR}}$ and the corresponding YLF zeros of a decoupled
RR with open boundary conditions undergoing the same quantum quench:
the blue dashed line and violet $\times$ symbols in Figs.~\hyperref[fig:YLF]{\ref{fig:YLF}(a)}
and \hyperref[fig:YLF]{\ref{fig:YLF}(b)}, respectively. We verify
that $f_{\text{RR}}(t)$ accurately reproduces the singular part of
$f(t)$, the difference being due to the analytical bulk's contribution.
Interestingly, we verify a one-to-one correspondence between the set
of zeros of $Z(z)$ near the real-time axis and the zeros of $Z_{\text{RR}}(z)$.
The differences between them vanish exponentially as $J_{\text{RR}}$
increases~\citep{supp-mat}.

We now further explore the consequences of our interpretation: (i)
Different RRs are independent (if sufficiently far from each other)
and (ii) the post-quench excitations are localized inside the RRs
(for sufficiently short times). The reasoning behind (i) is because
the bulk is practically in its ground state and thus its ground-state
correlation length $\xi$ is still a well-defined quantity.

\begin{figure}[b]
\begin{centering}
\includegraphics[clip,width=1\columnwidth]{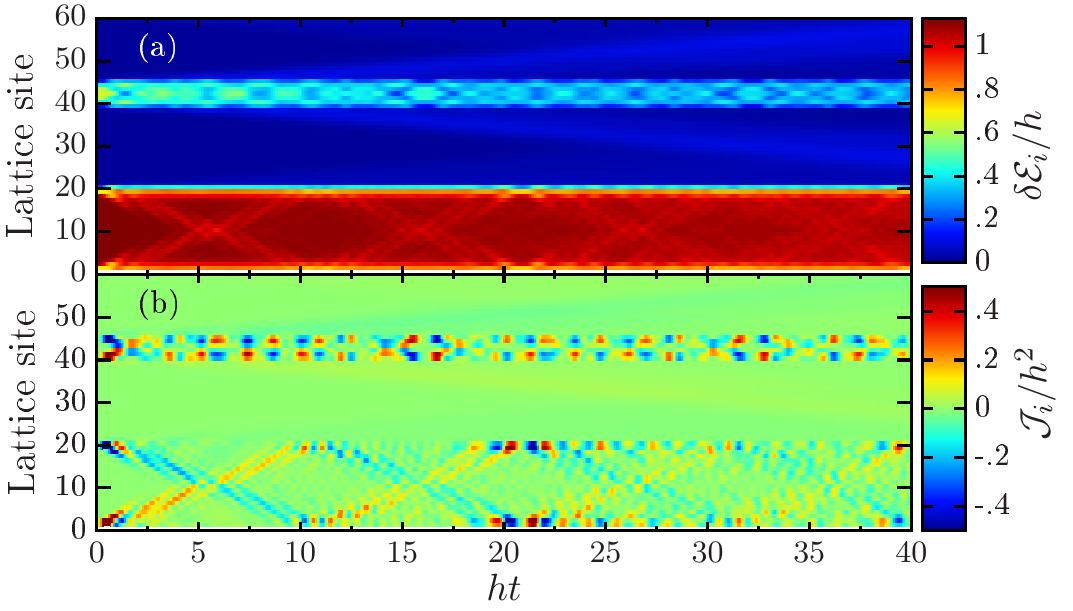}
\par\end{centering}
\caption{\label{fig:E-J-2RR}(a) The mean energy density above the ground state
$\delta{\cal E}_{i}$ and (b) the corresponding density current ${\cal J}_{i}$
as a function of the real time $t$ for each lattice site $i$ (see
text).}
\end{figure}

To give evidence of the above statements, we study the time evolution
of the mean energy density above the ground state $\delta{\cal E}_{i}=\left\langle \psi\left(t\right)\left|E_{i}\right|\psi\left(t\right)\right\rangle -\left\langle \phi_{\text{GS}}\left|E_{i}\right|\phi_{\text{GS}}\right\rangle $
(where $E_{i}=-\frac{1}{2}J_{i-1}\sigma_{i-1}^{z}\sigma_{i}^{z}-h\sigma_{i}^{x}-\frac{1}{2}J_{i}\sigma_{i}^{z}\sigma_{i+1}^{z}$
and $\left|\phi_{\text{GS}}\right\rangle $ is the ground state of
the post-quench $H$) and the associated density current ${\cal J}_{i}=hJ_{i-1}\left\langle \psi\left(t\right)\left|-\sigma_{i-1}^{z}\sigma_{i}^{y}+\sigma_{i-1}^{y}\sigma_{i}^{z}\right|\psi\left(t\right)\right\rangle $~\citep{supp-mat}.
We consider the same quench (from $h_{0}=\infty$ to $h$) in a chain
of $L=60$ sites long with periodic boundary conditions where the
bulk coupling is $J_{\text{B}}=0.2h$. The chain has two RRs. One
is $20$ sites long with coupling constant $J_{\text{RR},1}=2h$,
and the other is only $5$ sites long with couplings $J_{\text{RR},2}=1.5h$.
In Fig.~\ref{fig:E-J-2RR}, we plot the $\delta{\cal E}_{i}$ and
${\cal J}_{i}$ as a function of time.

Clearly, for the time window studied, the excitations are well localized
inside the RRs and the bulk remains in its ground state carrying no
energy current. We have also verified~\citep{supp-mat} that the
singular part of $f(t)$ and the corresponding YLF zeros are well
described by those of the same RRs undergoing the same quantum quench
but decoupled from the bulk.

Having demonstrated that (i) the RR dynamics is effectively decoupled
from the bulk and (ii) that the dynamics of sufficiently far apart
RRs are essentially independent from each other, we can readily understand
the origin and quantify the non-analyticities of $f(t)$ for any quantum
quench which does not cross the bulk QPT. All the singularities come
from sufficiently large RRs which, independently, provide YLF zeros
accumulating in lines piercing the real-time axis. Since the time
instant in which these lines pierce the real-time axis depends on
the microscopic details of the RRs, the YLF zeros will be generically
distributed over an area of the complex-time plane. The intersection
of this area with the real-time axis defines the dynamical quantum
Griffiths phase (see Fig.~\ref{fig:summary}).

Evidently, besides identifying the physical mechanism behind the non-analyticities
in $Z(t)$, it is also desirable to quantify it. From the Weierstrass
factorization theorem, the singular part of $f(t)$ is~\citep{yang-lee-pr52,fisher-book-1965,heyl-rpp18}
\begin{equation}
f_{\text{sing}}(t)\propto\sum_{m,\alpha}\ln\left|t-t_{m,\alpha}^{*}\right|\rightarrow\int\text{d}t^{*}g(t^{*})\ln\left|t-t^{*}\right|.\label{eq:fsing}
\end{equation}
 Here, $t_{m,\alpha}^{*}$ is the $m$th real-time YLF zero due to
the $\alpha$th RR. In the thermodynamic limit, the sum in Eq.~\eqref{eq:fsing}
is replaced by an integral weighted by the distribution of zeros $g(t^{*})$.
As noticed by Fisher~\citep{fisher-book-1965}, $f_{\text{sing}}$
is as a two-dimensional electrostatic potential due to point charges
at $t^{*}$. The non-analyticity of $f(t)$ is thus encoded in the
distribution $g(t)$, whose non-analyticities are inherited from the
distribution of the random variables in $H$. Naturally, an example
that can be worked out analytically is desirable. This is provided
by the percolating case in which the couplings are vanishing with
probability $p$ and equal to $J_{\alpha}>0$ with probability $1-p$.
Here, $J_{\alpha}$ is a random variable distributed according to
$P(J)$. For the quantum quench $h_{0}=\infty\rightarrow h=0_{+}$,
the dynamical free energy is~\citep{supp-mat} 
\begin{equation}
f=-L^{-1}\sum_{k=1}^{L}\ln\cos^{2}\left(J_{k}t\right)\rightarrow-\left(1-p\right)\overline{\ln\cos^{2}\left(Jt\right)},\label{eq:fsing-extreme}
\end{equation}
 where the thermodynamic limit was taken in the last passage and $\overline{(\cdots)}=\int\text{d}JP(J)(\cdots)$.
The real-time YLF zeros are $t_{m,\alpha}^{*}=\left(2m_{\alpha}+1\right)\pi/J_{\alpha}$.
If $J_{\alpha}$ is uniformly distributed between $J_{1}$ and $J_{2}$,
the non-analyticities of $P(J)$ at $J=J_{1(2)}$ become non-analyticities
of $f(t)$ at the time instants $t_{m,1(2)}^{c}=\left(2m_{1(2)}+1\right)\pi/J_{1(2)}$.
Notice that these are the only time instants in which $f(t)$ is non-analytic,
even though there is a continuum of YLF zeros in the time window $t_{m,2}^{c}<t<t_{m,1}^{c}$.
This is in close correspondence with the non-analyticities of the
electrostatic potential due to a continuous distribution of charges.
The associated singularities are only log-infinite derivatives of
$f$ at the instants $t_{m,2}^{c}$ and $t_{m,1}^{c}$. At all other
time instants, $f$ is locally analytic. At first glance, this seems
to imply a nearly undetectable non-analytical behavior (just as classical
Griffiths singularities). However, in numerical studies, the lack
of a dense accumulation of real-time zeros yields a highly fluctuating
free energy in that time window, as illustrated in Fig.~\hyperref[fig:summary]{\ref{fig:summary}(b)}.
Different convergence schemes or precisions will produce highly different
numerical results in the dynamical quantum Griffiths phase. We expect
an analogous behavior in the current experiments~\citep{heyl-etal-prl13,jucervic-etal-prl17,zhang-etal-nature17,bernie-etal-nature17,flaschner-etal-np18,guo-etal-pra19}
of ultracold atoms and other quantum simulators where the total number
of degrees of freedom is far from the thermodynamic limit. In electrostatics,
the same effect occurs if the probe of the electric field is able
to distinguish between neighboring point charges. Mathematically,
this is quantified by the Euler-Maclaurin formula of the difference
between the sum and integral in Eq.~\eqref{eq:fsing}, or, equivalently,
by the difference between the sample average (sum) and the distribution
average (integral) in Eq.~\eqref{eq:fsing-extreme}.

In summary, we have shown that RRs play a fundamental role in the
early-time dynamics of strongly interacting quantum systems after
quantum quenches which cross the RR QPT but not the bulk QPT. In that
case, the quench-induced excitations are confined in the RRs while
the bulk remains nearly in its ground state. As a result, observables
such as the dynamical free energy \eqref{eq:Z} become non-analytic
functions of time in the thermodynamic limit. The non-analyticities
are due to RR-induced YLF zeros accumulating in lines piercing the
real-time axis. Evidently, it is desirable to know whether this situation
applies to other model systems. For short times, we expect it to be
quite general when the bulk is gapped since there will be infrequent
resonances between the RRs and the bulk and thus the excitations remain
confined. For a gapless bulk, the RR relaxation time may still be
comparatively long since the nature of its excitations is fundamentally
different from the bulk's. In other words, the quench-induced excitations
in the RRs may not decay rapidly into the bulk due to the conservation
of emergent quantum numbers. We stress that, counterintuitively, the
RR-induced singular behavior of the dynamical free energy appears
at short timescales. This fact makes the RR-induced singularities
easier to be identified in numerical studies (such as time-dependent
density matrix renormalization group) and in quantum simulator experiments
(before the interactions with the environment spoil the unitary dynamics). 

Notice that the non-equilibrium phenomenon here studied is of short-time
scales. Studying (the long-time physics of) thermalization after the
quantum quenches here considered (when integrability-breaking terms
are present) by quantifying how the excitations decay into the bulk
and relating this to the position of the YLF zeros is an interesting
task left for the future.

Finally, we remark that our results also apply to quantum annealing~\citep{gardas-etal-sr18}
from $h_{0}$ to $h$ when the RR QPT is crossed. If the RR is sufficiently
large or the annealing is sufficiently fast, excitations are generated
and confined inside the RR. Thus, RRs play an important role for adiabatic
quantum computing.
\begin{acknowledgments}
We acknowledge instructive discussions with Markus Heyl, David Luitz,
Roderich Moessner, and Matthias Vojta. We also acknowledge the financial
support of the Brazilian agencies FAPEMIG, FAPESP, and CNPq. 
\end{acknowledgments}

\newpage{}

%%%%%%%%%% Merge with supplemental materials %%%%%%%%%% 
\clearpage
\widetext
%%%%%%%%%% Prefix a "S" to all equations, figures, tables and reset the counter %%%%%%%%%% 
\setcounter{equation}{0} 
\setcounter{figure}{0} 
\setcounter{table}{0} 
\setcounter{page}{1} 
\setcounter{section}{0}
\makeatletter 
\renewcommand{\thetable}{S\arabic{table}}
\renewcommand{\theequation}{S\arabic{equation}} 
\renewcommand{\thefigure}{S\arabic{figure}} 
%\renewcommand{\bibnumfmt}[1]{[S#1]} 
%\renewcommand{\citenumfont}[1]{S#1} 
%%%%%%%%%% Prefix a "S" to all equations, figures, tables and reset the counter %%%%%%%%%%
\begin{center}
\textbf{\large{}Supplementary Material for ``Disorder-induced dynamical
Griffiths singularities after certain quantum quenches''}{\large\par}
\par\end{center}

\begin{center}
Jos\'e A. Hoyos,$^{1,2}$ R. F. P. Costa,$^{3}$ and J. C. Xavier$^{3}$\vspace{-0.2cm}
\par\end{center}

\begin{center}
$^{1}$\emph{\small{}Instituto de F\'{\i}sica de S\~ao Carlos, Universidade
de S\~ao Paulo, C.~P.~369, S\~ao Carlos, S\~ao Paulo 13560-970,
Brazil}\\
$^{2}$\emph{\small{}Max Planck Institute for the Physics of Complex
Systems, N\"othnitzer Str. 38, 01187 Dresden, Germany}\\
$^{3}$\emph{\small{}Universidade Federal de Uberl\^andia, Instituto
de F\'{\i}sica, C.~P.~593, 38400-902 Uberl\^andia, MG, Brazil}{\small\par}
\par\end{center}

\section{The zero-temperature phase diagram\label{sec:SM}}

\begin{figure}[h]
\begin{centering}
\includegraphics[clip,width=7cm]{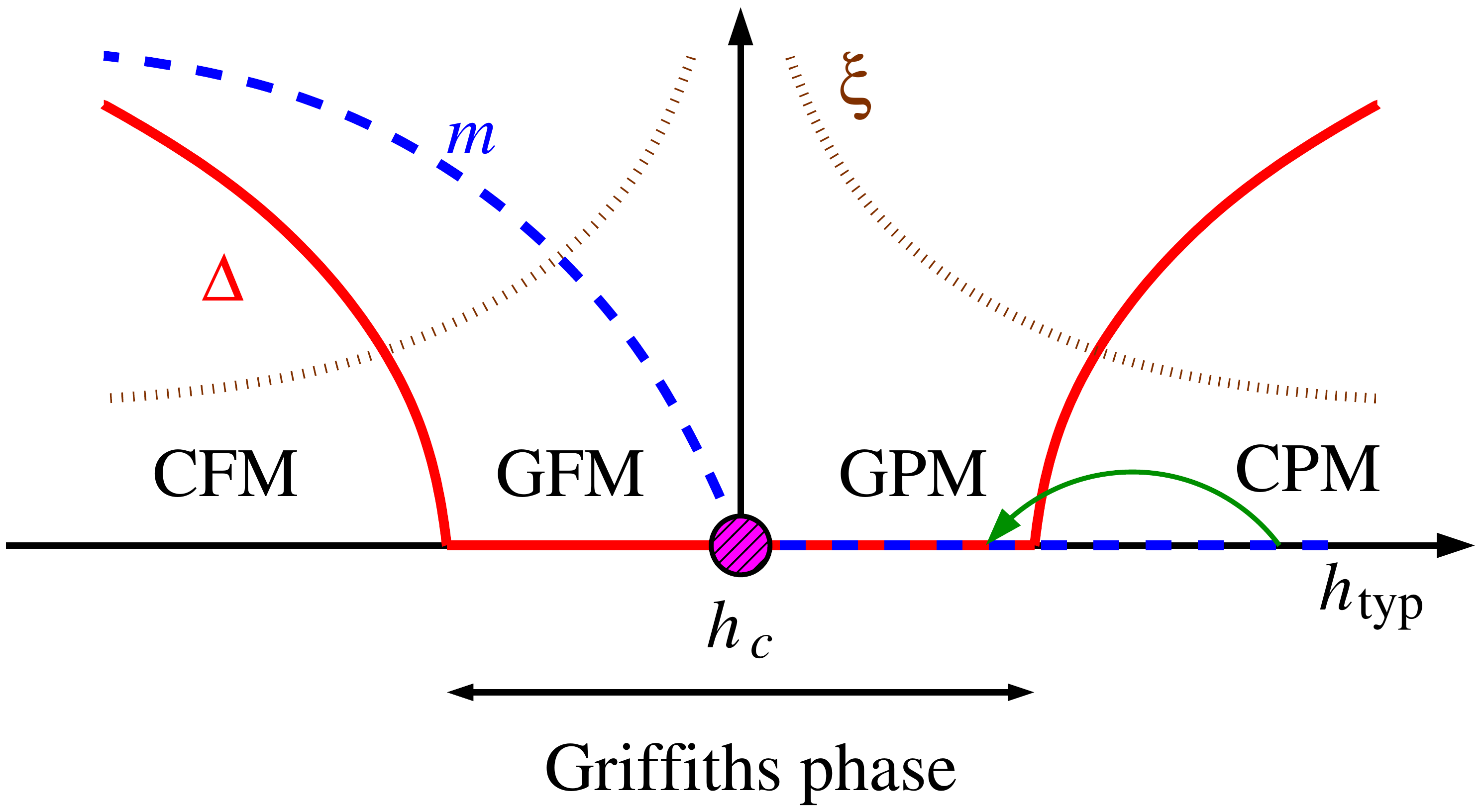}
\par\end{centering}
\caption{Schematics of the zero-temperature phase diagram of the model Hamiltonian
\eqref{eq:H} of the main text or, more generically, Eq.~\eqref{eq:H-free-fermion},
the corresponding spectral gap $\Delta$ (red), the order parameter
$m$ (dahsed blue), and the spin-spin correlation length $\xi$ (dotted
brown). The critical point at $h_{\text{typ}}=h_{c}=J_{\text{typ}}$
is surrounded by the ferromagnetic and paramagnetic Griffiths phases
(GFM and GPM, respectively) where $\Delta$ vanishes. Here, CFM and
CPM stand for conventional ferromagnetic and paramagnetic phases,
respectively. The quantum quenches we study in this work are depicted
by the (green) arrow connecting the CPM to the GPM phases.\label{fig:PD}}
\end{figure}

The effects of random disorder on the zero-temperature phase diagram
of the transverse-field Ising chain, Eq.~\eqref{eq:H} of the main
text or, more generally, Eq.~\eqref{eq:H-Ising}, is well understood.
The critical point (of infinite-randomness type~\citep{fisher95})
takes place when the typical values of the odd and even couplings
are equal~\citep{pfeuty-pla79}, i.e., when 
\[
h_{\text{typ}}\equiv\exp\left(\overline{\ln h_{i}}\right)=h_{c}=J_{\text{typ}}\equiv\exp\left(\overline{\ln J_{i}}\right),
\]
 where $\overline{\cdots}$ denotes the disorder average. Surrounding
the critical point, there are the paramagnetic and the ferromagnetic
Griffiths phases. These phases have the same nature of their clean
counterparts in the sense that the order parameter $m=\overline{\left\langle \sigma_{i}^{z}\right\rangle }$
is finite (vanishing) in the ferromagnetic (paramagnetic) phase, and
the spin-spin correlation length $\xi$ is finite. However, the gap
$\Delta$ in the energy spectrum vanishes throughout these Griffiths
phases~\citep{fisher95}. Schematically, the phase diagram, the order
parameter $m$, the excitation gap $\Delta$ are shown in Fig.~\ref{fig:PD}. 

The extent of the Griffiths phase is proportional to the disorder
strength of the coupling constants. For concreteness, let $\{J_{i}\}$
be independent random variables distributed between $J_{\text{min}}<J_{i}<J_{\text{max}}$.
The Griffiths paramagnetic phase covers the interval $h_{c}<h_{\text{typ}}<J_{\text{max}}$
and the Griffiths ferromagnetic phase covers the interval $J_{\text{min}}<h_{\text{typ}}<h_{c}$. 

Deep in the conventional phases, the ground state is very similar
to the clean one, and the system properties (like the spectral gap)
are well approximated by that of the clean system with the value of
the clean $J$ and $h$ being replaced by its typical values. 

In this work, we show the relevance of the rare regions on the unitary
dynamics after a quantum quench from the conventional paramagnetic
phase to the nearby Griffiths paramagnetic phase (see green arrow
in Fig.~\ref{fig:PD}). In this quench, the bulk experiences a ``mild''
quench and, thus, remains nearly in its ground state. On the other
hand, this quantum quench brings the rare regions from one phase to
the other, and, thus, are highly excited. 

\section{Mapping to free fermions, diagonalization, and observables}

\subsection{The mapping}

Following Refs.~\citealp{lieb-schultz-mattis,young-rieger-prb96},
the random transverse-field Ising chain Hamiltonian with periodic
boundary conditions is 
\begin{equation}
H=-\sum_{i=1}^{L}J_{i}\sigma_{i}^{z}\sigma_{i+1}^{z}-\sum_{i=1}^{L}h_{i}\sigma_{i}^{x},\label{eq:H-Ising}
\end{equation}
 which generalizes the Hamiltonian \eqref{eq:H} of the main text,
can be mapped to a free femionic one via the Jordan-Wigner transformation
\begin{equation}
\sigma_{j}^{x}=1-2n_{j}=c_{j}^{\phantom{\dagger}}c_{j}^{\dagger}-c_{j}^{\dagger}c_{j}^{\phantom{\dagger}},\quad\sigma_{j}^{y}=ie^{i\pi\sum_{k=1}^{j-1}n_{k}}\left(c_{j}^{\dagger}-c_{j}^{\phantom{\dagger}}\right),\quad\sigma_{j}^{z}=e^{i\pi\sum_{k=1}^{j-1}n_{k}}\left(c_{j}^{\dagger}+c_{j}^{\phantom{\dagger}}\right),\label{eq:mapping}
\end{equation}
 with $\left\{ c_{i}\right\} $ being fermionic operators of spinless
fermions, i.e., $\left\{ c_{i}^{\dagger},c_{j}^{\dagger}\right\} =\left\{ c_{i}^{\phantom{\dagger}},c_{j}^{\phantom{\dagger}}\right\} =0$
and $\left\{ c_{i}^{\phantom{\dagger}},c_{j}^{\dagger}\right\} =\delta_{i,j}$.
The corresponding fermionic Hamiltonian is 
\begin{equation}
H=\left(\mathbf{c}^{\dagger}\right)^{T}\mathbb{A}\mathbf{c}-\left(\mathbf{c}\right)^{T}\mathbb{A}\mathbf{c}^{\dagger}+\left(\mathbf{c}^{\dagger}\right)^{T}\mathbb{B}\mathbf{c}^{\dagger}-\left(\mathbf{c}\right)^{T}\mathbb{B}\mathbf{c},\label{eq:H-free-fermion}
\end{equation}
 where $\mathbf{c}^{T}=\left(c_{1}^{\phantom{\dagger}},c_{L}^{\phantom{\dagger}},\dots,c_{L}^{\phantom{\dagger}}\right)$
is a row vector operator and the matrices $\mathbb{A}$ and $\mathbb{B}$
are 
\begin{equation}
\mathbb{A}=\frac{1}{2}\left(\begin{array}{ccccc}
2h_{1} & -J_{1} & 0 & \cdots & \left(-1\right)^{N}J_{L}\\
-J_{1} & 2h_{2} & -J_{2} & \cdots & 0\\
0 & -J_{2} & 2h_{3} & \cdots & \vdots\\
\vdots & \vdots & \vdots & \ddots & -J_{L-1}\\
\left(-1\right)^{N}J_{L} & 0 & \cdots & -J_{L-1} & 2h_{L}
\end{array}\right),\ \mathbb{B}=\frac{1}{2}\left(\begin{array}{ccccc}
0 & -J_{1} & 0 & \cdots & -\left(-1\right)^{N}J_{L}\\
J_{1} & 0 & -J_{2} & \cdots & 0\\
0 & J_{2} & 0 & \cdots & \vdots\\
\vdots & \vdots & \vdots & \ddots & -J_{L-1}\\
\left(-1\right)^{N}J_{L} & 0 & \cdots & J_{L-1} & 0
\end{array}\right).
\end{equation}
 Here, $N=\sum_{j=1}^{2L}n_{j}$ is the total number of fermions.
Although $N$ it is not a conserved quantity, its parity is. Thus,
$e^{i\pi N}=\left(-1\right)^{N}$ is a conserved quantity. The value
of the parity is determined by that one giving the lowest ground-state
energy.

\subsection{Diagonalization}

The diagonalization is via the Bogoliubov-Valatin transformation~\citep{young-rieger-prb96}.
Thus, defining the matrix $\mathbb{M}$ such that 
\begin{equation}
H=\left(\begin{array}{cc}
\mathbf{c}^{\dagger T} & \mathbf{c}^{T}\end{array}\right)\mathbb{M}\left(\begin{array}{c}
\mathbf{c}\\
\mathbf{c}^{\dagger}
\end{array}\right),\mbox{ then }\mathbb{M}=\left(\begin{array}{cc}
\mathbb{A} & \mathbb{B}\\
-\mathbb{B} & -\mathbb{A}
\end{array}\right),
\end{equation}
 is a $2L\times2L$ symmetric matrix. The matrix $\mathbb{M}$ is
brought to a diagonal form $\mathbb{D}=\mathbb{V^{T}MV}$, with $\mathbb{V}$
being a matrix whose the $k$th column is the $k$th eigenvectors
of $\mathbb{M}$ and $\mathbb{D}=\text{diag}\left\{ \lambda_{k}\right\} $
is a diagonal matrix whose elements are the corresponding eigenenergies.
It is possible to show that the eigenenergies appears in positive-negative
pairs, i.e., if $\lambda>0$ is an eigenenergy of $\mathbb{M}$, so
is $-\lambda$. In addition, it is possible to show that $\mathbb{V}$
can be written as 
\begin{equation}
\mathbb{V}=\left(\begin{array}{cc}
\mathbb{R} & \mathbb{L}\\
\mathbb{L} & \mathbb{R}
\end{array}\right),\label{eq:Vmat}
\end{equation}
 where $\mathbb{R}^{T}\mathbb{R}+\mathbb{L}^{T}\mathbb{L}=\mathbb{R}\mathbb{R}^{T}+\mathbb{L}\mathbb{L}^{T}=\mathds{1}$
and $\mathbb{R}^{T}\mathbb{L}+\mathbb{L}^{T}\mathbb{R}=\mathbb{L}\mathbb{R}^{T}+\mathbb{R}\mathbb{L}^{T}=\mathbf{0}$.
In addition, the first $L$ eigenenergies are $\lambda_{k}\ge0$ while
the remaining ones are negative with $\lambda_{k+L}=-\lambda_{k}$.

With these properties, the Hamiltonian can be brought to a diagonal
form 
\begin{equation}
H=\sum_{k=1}^{L}\lambda_{k}\left(\gamma_{k}^{\dagger}\gamma_{k}^{\phantom{\dagger}}-\gamma_{k}^{\phantom{\dagger}}\gamma_{k}^{\dagger}\right),
\end{equation}
 where all eigenenergies $\left\{ \lambda_{k}\right\} $, $k=1,\dots,L$
are non-negative, and $\left\{ \gamma_{k}\right\} $ are fermionic
operators which are related to the original fermions via 
\begin{equation}
\mathbf{c}=\mathbb{R}\boldsymbol{\gamma}+\mathbb{L}\boldsymbol{\gamma}^{\dagger},\mbox{ and }\boldsymbol{\gamma}=\mathbb{R}^{T}\mathbf{c}+\mathbb{L}^{T}\mathbf{c}^{\dagger}.\label{eq:c-gamma}
\end{equation}

In order to determine the parity $\left(-1\right)^{N}$ of the ground
state, we need, in general, to diagonalize $\mathbb{M}$ with both
parities and pick up the one yielding the lowest ground state energy
\begin{equation}
E_{\text{GS}}=-\sum_{k=1}^{L}\lambda_{k}.\label{eq:EGS}
\end{equation}
 Once the parity is determined for the pre-quench Hamiltonian, it
is conserved by the post-quench Hamiltonian.

\subsection{The dynamical partition function}

We now want to compute the return probability amplitude {[}Eq.~\eqref{eq:Z}
of the main text{]} 
\begin{equation}
Z\left(z\right)=\left\langle \psi_{0}\left|e^{-iHz}\right|\psi_{0}\right\rangle \equiv\left\langle e^{-iHz}\right\rangle _{0},
\end{equation}
 where $\left|\psi_{0}\right\rangle $ is the ground state of the
pre-quench Hamiltonian $H_{0}$, $H$ is the post-quench Hamiltonian,
and $z=t+i\tau$ is the complex time. Evidently, we are assuming that
$H_{0}$ and $H$ can be written as free-fermionic Hamiltonians \eqref{eq:H-free-fermion}.

Following \citep{lieb-schultz-mattis}, we recast 
\begin{eqnarray}
e^{-iHz} & = & \prod_{k=1}^{L}e^{-i\lambda_{k}z\left(\gamma_{k}^{\dagger}\gamma_{k}^{\phantom{\dagger}}-\gamma_{k}^{\phantom{\dagger}}\gamma_{k}^{\dagger}\right)}=e^{-iE_{\text{GS}}z}\prod_{k=1}^{L}e^{-i2\lambda_{k}z\left(\gamma_{k}^{\dagger}\gamma_{k}^{\phantom{\dagger}}\right)}\nonumber \\
 & = & e^{-iE_{\text{GS}}z}\prod_{k=1}^{L}\left(e^{-2i\lambda_{k}z}\gamma_{k}^{\dagger}+\gamma_{k}^{\phantom{\dagger}}\right)\left(\gamma_{k}^{\dagger}+\gamma_{k}^{\phantom{\dagger}}\right)=e^{-iE_{\text{GS}}z}\prod_{k=1}^{L}A_{k}B_{k},\label{eq:AB-G}
\end{eqnarray}
 where the vector operators 
\begin{eqnarray}
\mathbf{A} & = & e^{-2i\boldsymbol{\lambda}z}\boldsymbol{\gamma}^{\dagger}+\boldsymbol{\gamma}=\left(\mathbb{L}^{T}+e^{-2i\boldsymbol{\lambda}z}\mathbb{R}^{T}\right)\mathbf{c}^{\dagger}+\left(e^{-2i\boldsymbol{\lambda}z}\mathbb{L}^{T}+\mathbb{R}^{T}\right)\mathbf{c}\nonumber \\
 & = & \left(\mathbb{L}^{T}+e^{-2i\boldsymbol{\lambda}z}\mathbb{R}^{T}\right)\left(\mathbb{R}_{0}\boldsymbol{\gamma}_{0}^{\dagger}+\mathbb{L}_{0}\boldsymbol{\gamma}_{0}^{\phantom{\dagger}}\right)+\left(e^{-2i\boldsymbol{\lambda}z}\mathbb{L}^{T}+\mathbb{R}^{T}\right)\left(\mathbb{R}_{0}\boldsymbol{\gamma}_{0}^{\phantom{\dagger}}+\mathbb{L}_{0}\boldsymbol{\gamma}_{0}^{\dagger}\right),\\
\mathbf{B} & = & \boldsymbol{\gamma}^{\dagger}+\boldsymbol{\gamma}=\left(\mathbb{L}^{T}+\mathbb{R}^{T}\right)\left(\mathbb{L}_{0}+\mathbb{R}_{0}\right)\left(\boldsymbol{\gamma}_{0}^{\dagger}+\boldsymbol{\gamma}_{0}^{\phantom{\dagger}}\right).
\end{eqnarray}
 Here, $e^{-2i\boldsymbol{\lambda}z}$ is a $L\times L$ diagonal
matrix with the $k$th diagonal element $e^{-2i\lambda_{k}z}$, and
$\boldsymbol{\gamma}^{T}=\left(\gamma_{1},\dots,\gamma_{L}\right)$
with $\left\{ \lambda_{k}\right\} $ being the eigenfermions of the
post-quench Hamiltonian. Likewise, $\boldsymbol{\gamma}_{0}^{T}$
is the analog for $H_{0}$.

The mean value $\left\langle e^{-iHz}\right\rangle _{0}$ is obtained
by the use of the Wick's theorem. We then need all non-vanishing contractions
of $\prod_{k=1}^{L}A_{k}B_{k}$. The contractions of type $\left\langle A_{j}A_{k}\right\rangle $
do not give a diagonal matrix, but the contraction of the $B$'s does:
\[
\left\langle \mathbf{B}\mathbf{B}^{T}\right\rangle =\left(\mathbb{L}^{T}+\mathbb{R}^{T}\right)\left(\mathbb{L}_{0}+\mathbb{R}_{0}\right)\left\langle \left(\boldsymbol{\gamma}_{0}^{\dagger}+\boldsymbol{\gamma}_{0}^{\phantom{\dagger}}\right)\left(\boldsymbol{\gamma}_{0}^{\dagger T}+\boldsymbol{\gamma}_{0}^{T}\right)\right\rangle _{0}\left(\mathbb{L}_{0}^{T}+\mathbb{R}_{0}^{T}\right)\left(\mathbb{L}+\mathbb{R}\right)=\mathds{1},
\]
 since $\left\langle \boldsymbol{\gamma}_{0}^{\dagger}\boldsymbol{\gamma}_{0}^{\dagger T}\right\rangle _{0}=\left\langle \boldsymbol{\gamma}_{0}^{\dagger}\boldsymbol{\gamma}_{0}^{T}\right\rangle _{0}=\left\langle \boldsymbol{\gamma}_{0}^{\phantom{\dagger}}\boldsymbol{\gamma}_{0}^{T}\right\rangle _{0}=\mathbf{0}$
and $\left\langle \boldsymbol{\gamma}_{0}^{\phantom{\dagger}}\boldsymbol{\gamma}_{0}^{\dagger T}\right\rangle _{0}=\mathds{1}$,
as there is no eigenfermion in the ground state of $H_{0}$. Therefore,
all contractions must be of type $\left\langle A_{i}B_{j}\right\rangle $.
A contraction of type $\left\langle A_{i}A_{j}\right\rangle $ is
not necessarily vanishing, however, it must be multiplied by a contraction
of type $\left\langle B_{m}B_{n}\right\rangle $ which vanishes since
$m\neq n$. Finally, we have that 
\begin{equation}
\left\langle e^{-iHz}\right\rangle _{0}=e^{-iE_{\text{GS}}z}\left\langle \prod_{k=1}^{L}A_{k}B_{k}\right\rangle _{0}=e^{-iE_{\text{GS}}z}\det\left(\begin{array}{cccc}
\left\langle A_{1}B_{1}\right\rangle _{0} & \left\langle A_{1}B_{2}\right\rangle _{0} & \cdots & \left\langle A_{1}B_{L}\right\rangle _{0}\\
\left\langle A_{2}B_{1}\right\rangle _{0} & \left\langle A_{2}B_{2}\right\rangle _{0} & \cdots & \left\langle A_{2}B_{L}\right\rangle _{0}\\
\vdots & \vdots & \ddots & \vdots\\
\left\langle A_{L}B_{1}\right\rangle _{0} & \left\langle A_{L}B_{2}\right\rangle _{0} & \cdots & \left\langle A_{L}B_{L}\right\rangle _{0}
\end{array}\right)=e^{-iE_{\text{GS}}z}\det\left\langle \mathbf{A}\mathbf{B}^{T}\right\rangle _{0}.\label{eq:Det-G}
\end{equation}
 The mean value 
\begin{eqnarray*}
\left\langle \mathbf{A}\mathbf{B}^{T}\right\rangle _{0} & = & \left[\left(\mathbb{L}^{T}+e^{-2i\boldsymbol{\lambda}z}\mathbb{R}^{T}\right)\mathbb{L}_{0}+\left(e^{-2i\boldsymbol{\lambda}z}\mathbb{L}^{T}+\mathbb{R}^{T}\right)\mathbb{R}_{0}\right]\left\langle \boldsymbol{\gamma}_{0}^{\phantom{\dagger}}\boldsymbol{\gamma}_{0}^{\dagger}\right\rangle _{0}\left(\mathbb{L}_{0}^{T}+\mathbb{R}_{0}^{T}\right)\left(\mathbb{L}+\mathbb{R}\right)\\
 & = & \left[e^{-2i\boldsymbol{\lambda}z}\left(\mathbb{R}^{T}\mathbb{L}_{0}+\mathbb{L}^{T}\mathbb{R}_{0}\right)+\mathbb{L}^{T}\mathbb{L}_{0}+\mathbb{R}^{T}\mathbb{R}_{0}\right]\left(\mathbb{L}_{0}^{T}+\mathbb{R}_{0}^{T}\right)\left(\mathbb{L}+\mathbb{R}\right).
\end{eqnarray*}
 Summarizing, 
\begin{equation}
Z\left(z\right)=e^{-iE_{\text{GS}}z}\det\left[\left(e^{-2i\boldsymbol{\lambda}z}-\mathds{1}\right)\left(\mathbb{R}^{T}\mathbb{L}_{0}+\mathbb{L}^{T}\mathbb{R}_{0}\right)\left(\mathbb{L}_{0}^{T}+\mathbb{R}_{0}^{T}\right)\left(\mathbb{L}+\mathbb{R}\right)+\mathds{1}\right].\label{eq:AB}
\end{equation}
 We checked this result against exact diagonalization of \eqref{eq:H-Ising}
in the spin basis for various quantum quenches, complex time instants
$z$, coupling configurations $\left\{ J_{i}\right\} $, and chain
sizes from $L=4$ to $8$. The difference is within machine precision.

\subsection{Energy density and current}

The energy current operator is obtained in the following manner. Define
the energy density operator as 
\begin{equation}
E_{i}=-\frac{1}{2}J_{i-1}\sigma_{i-1}^{z}\sigma_{i}^{z}-h_{i}\sigma_{i}^{x}-\frac{1}{2}J_{i}\sigma_{i}^{z}\sigma_{i+1}^{z}.\label{eq:local-E}
\end{equation}
 The Hamiltonian is $H=\sum_{i}E_{i}$ and is a conserved quantity.
Thus, there is a continuity equation 
\[
\frac{\partial{\cal E}_{i}}{\partial t}+\nabla\cdot{\cal J}=\frac{\partial{\cal E}_{i}}{\partial t}+\left({\cal J}_{i+1}-{\cal J}_{i}\right)=0,
\]
 where ${\cal E}_{i}=\left\langle \psi\left(t\right)\left|E_{i}\right|\psi\left(t\right)\right\rangle $,
${\cal J}_{i}=\left\langle \psi\left(t\right)\left|I_{i}\right|\psi\left(t\right)\right\rangle $
is the mean value of the associated local energy current operator,
and we have taken the discrete divergent considering the lattice spacing
as 1. The energy current operator is obtained from the continuity
equation 
\[
\frac{\partial{\cal E}_{i}}{\partial t}=i\left\langle \left[H,E_{i}\right]\right\rangle .
\]
 The commutator is simply 
\begin{eqnarray*}
\left[H,E_{i}\right] & = & i\left[J_{i}\left(-h_{i+1}\sigma_{i}^{z}\sigma_{i+1}^{y}+h_{i}\sigma_{i}^{y}\sigma_{i+1}^{z}\right)-J_{i-1}\left(-h_{i}\sigma_{i-1}^{z}\sigma_{i}^{y}+h_{i-1}\sigma_{i-1}^{y}\sigma_{i}^{z}\right)\right]=i\left(I_{i+1}-I_{i}\right),
\end{eqnarray*}
i.e., the current operator is 
\begin{equation}
I_{j}=J_{j-1}\left(-h_{j}\sigma_{j-1}^{z}\sigma_{j}^{y}+h_{j-1}\sigma_{j-1}^{y}\sigma_{j}^{z}\right).\label{eq:current-operator}
\end{equation}
 This recovers the current operator quoted in the main text. In the
free-fermionic language \eqref{eq:mapping}, 
\begin{eqnarray}
E_{j} & = & -\frac{1}{2}J_{j-1}\left(c_{j-1}^{\dagger}-c_{j-1}^{\phantom{\dagger}}\right)\left(c_{j}^{\dagger}+c_{j}^{\phantom{\dagger}}\right)+h_{j}\left(c_{j}^{\dagger}-c_{j}^{\phantom{\dagger}}\right)\left(c_{j}^{\dagger}+c_{j}^{\phantom{\dagger}}\right)-\frac{1}{2}J_{j}\left(c_{j}^{\dagger}-c_{j}^{\phantom{\dagger}}\right)\left(c_{j+1}^{\dagger}+c_{j+1}^{\phantom{\dagger}}\right),\label{eq:E-density-i}\\
I_{j} & = & iJ_{j-1}\left[-h_{j}\left(c_{j-1}^{\dagger}-c_{j-1}^{\phantom{\dagger}}\right)\left(c_{j}^{\dagger}-c_{j}^{\phantom{\dagger}}\right)+h_{j-1}\left(c_{j-1}^{\dagger}+c_{j-1}^{\phantom{\dagger}}\right)\left(c_{j}^{\dagger}+c_{j}^{\phantom{\dagger}}\right)\right].\label{eq:current-i}
\end{eqnarray}
 For the boundary terms, one simply replaces $j-1=0\rightarrow L$,
$j+1=L+1\rightarrow1$, and multiplies the resulting term by $\left(-1\right)^{N+1}$.

The average values of $\left(c_{j-1}^{\dagger}-c_{j-1}^{\phantom{\dagger}}\right)\left(c_{j}^{\dagger}+c_{j}^{\phantom{\dagger}}\right)$,
$\left(c_{j-1}^{\dagger}-c_{j-1}^{\phantom{\dagger}}\right)\left(c_{j}^{\dagger}-c_{j}^{\phantom{\dagger}}\right)$
and $\left(c_{j-1}^{\dagger}+c_{j-1}^{\phantom{\dagger}}\right)\left(c_{j}^{\dagger}+c_{j}^{\phantom{\dagger}}\right)$
(and the corresponding boundary terms) are needed. They all can be
obtained in the following unified way. Let $x=\pm1$ and $y=\pm1$,
then 
\begin{eqnarray}
\left\langle \left(\mathbf{c}^{\dagger}+x\mathbf{c}\right)\left(\mathbf{c}^{\dagger T}+y\mathbf{c}^{T}\right)\right\rangle  & = & \mathbb{P}_{x}\left(ye^{2i\boldsymbol{\lambda}t}\left\langle \boldsymbol{\gamma}^{\dagger}\boldsymbol{\gamma}^{T}\right\rangle _{0}e^{-2i\boldsymbol{\lambda}t}+e^{2i\boldsymbol{\lambda}t}\left\langle \boldsymbol{\gamma}^{\dagger}\boldsymbol{\gamma}^{\dagger T}\right\rangle _{0}e^{2i\boldsymbol{\lambda}t}\right.\nonumber \\
 &  & \left.+xye^{-2i\boldsymbol{\lambda}t}\left\langle \boldsymbol{\gamma}\boldsymbol{\gamma}^{T}\right\rangle _{0}e^{-2i\boldsymbol{\lambda}t}+xe^{-2i\boldsymbol{\lambda}t}\left\langle \boldsymbol{\gamma}\boldsymbol{\gamma}^{\dagger T}\right\rangle _{0}e^{2i\boldsymbol{\lambda}t}\right)\mathbb{P}_{y}^{T},\label{eq:cxccyc}
\end{eqnarray}
 where $\mathbb{P}_{x}=\mathbb{R}+x\mathbb{L}$ and we have used Eq.~\eqref{eq:c-gamma}.
It is a tedious algebra to show that 
\begin{eqnarray}
\left\langle \boldsymbol{\gamma}^{\dagger}\boldsymbol{\gamma}^{\dagger T}\right\rangle _{0} & = & \mathbb{R}^{T}\mathbb{L}_{0}\mathbb{R}_{0}^{T}\mathbb{R}+\mathbb{R}^{T}\mathbb{L}_{0}\mathbb{L}_{0}^{T}\mathbb{L}+\mathbb{L}^{T}\mathbb{R}_{0}\mathbb{R}_{0}^{T}\mathbb{R}+\mathbb{L}^{T}\mathbb{R}_{0}\mathbb{L}_{0}\mathbb{L},\label{eq:g+g+}\\
\left\langle \boldsymbol{\gamma}^{\dagger}\boldsymbol{\gamma}^{T}\right\rangle _{0} & = & \mathbb{R}^{T}\mathbb{L}_{0}\mathbb{L}_{0}^{T}\mathbb{R}+\mathbb{R}^{T}\mathbb{L}_{0}\mathbb{R}_{0}^{T}\mathbb{L}+\mathbb{L}^{T}\mathbb{R}_{0}\mathbb{L}_{0}^{T}\mathbb{R}+\mathbb{L}^{T}\mathbb{R}_{0}\mathbb{R}_{0}^{T}\mathbb{L},\label{eq:g+g}\\
\left\langle \boldsymbol{\gamma}\boldsymbol{\gamma}^{\dagger T}\right\rangle _{0} & = & \mathbb{R}^{T}\mathbb{R}_{0}\mathbb{R}_{0}^{T}\mathbb{R}+\mathbb{R}^{T}\mathbb{R}_{0}\mathbb{L}_{0}^{T}\mathbb{L}+\mathbb{L}^{T}\mathbb{L}_{0}\mathbb{R}_{0}^{T}\mathbb{R}+\mathbb{L}^{T}\mathbb{L}_{0}\mathbb{L}_{0}^{T}\mathbb{L},\label{eq:gg+}\\
\left\langle \boldsymbol{\gamma}\boldsymbol{\gamma}^{T}\right\rangle _{0} & = & \mathbb{R}^{T}\mathbb{R}_{0}\mathbb{L}_{0}^{T}\mathbb{R}+\mathbb{R}^{T}\mathbb{R}_{0}\mathbb{R}_{0}^{T}\mathbb{L}+\mathbb{L}^{T}\mathbb{L}_{0}\mathbb{L}_{0}^{T}\mathbb{R}+\mathbb{L}^{T}\mathbb{L}_{0}\mathbb{R}_{0}^{T}\mathbb{L}.\label{eq:gg}
\end{eqnarray}
 It is curious to notice that $\left\langle \boldsymbol{\gamma}\boldsymbol{\gamma}^{T}\right\rangle _{0}^{T}=\left\langle \boldsymbol{\gamma}^{\dagger}\boldsymbol{\gamma}^{\dagger T}\right\rangle _{0},\mbox{ and that }\left\langle \boldsymbol{\gamma}\boldsymbol{\gamma}^{\dagger T}\right\rangle _{0}^{T}+\left\langle \boldsymbol{\gamma}^{\dagger}\boldsymbol{\gamma}^{T}\right\rangle _{0}=\mathds{1}.$ 

Plugging \eqref{eq:g+g+}\textendash \eqref{eq:gg} into \eqref{eq:cxccyc},
we then find that 
\begin{align}
\left\langle \left(c_{j-1}^{\dagger}+c_{j-1}^{\phantom{\dagger}}\right)\left(c_{j}^{\dagger}+c_{j}^{\phantom{\dagger}}\right)\right\rangle \nonumber \\
=i\sum_{k,l=1}^{L}\mathbb{P}_{j-1,k} & \left(\cos\left(2\lambda_{k}t\right)\left(\mathbb{P}^{T}\mathbb{P}_{0}\mathbb{Q}_{0}^{T}\mathbb{Q}\right)_{k,l}\sin\left(2\lambda_{l}t\right)-\sin\left(2\lambda_{k}t\right)\left(\mathbb{P}^{T}\mathbb{P}_{0}\mathbb{Q}_{0}^{T}\mathbb{Q}\right)_{l,k}\cos\left(2\lambda_{l}t\right)\right)\mathbb{P}_{l,j}^{T},\\
\left\langle \left(c_{j-1}^{\dagger}-c_{j-1}^{\phantom{\dagger}}\right)\left(c_{j}^{\dagger}-c_{j}^{\phantom{\dagger}}\right)\right\rangle \nonumber \\
=i\sum_{k,l=1}^{L}\mathbb{Q}_{j-1,k} & \left(\sin\left(2\lambda_{k}t\right)\left(\mathbb{P}^{T}\mathbb{P}_{0}\mathbb{Q}_{0}^{T}\mathbb{Q}\right)_{k,l}\cos\left(2\lambda_{l}t\right)-\cos\left(2\lambda_{k}t\right)\left(\mathbb{P}^{T}\mathbb{P}_{0}\mathbb{Q}_{0}^{T}\mathbb{Q}\right)_{k,l}\sin\left(2\lambda_{l}t\right)\right)\mathbb{Q}_{l,j}^{T},\\
\left\langle \left(c_{j-1}^{\dagger}-c_{j-1}^{\phantom{\dagger}}\right)\left(c_{j}^{\dagger}+c_{j}^{\phantom{\dagger}}\right)\right\rangle \nonumber \\
=-\sum_{k,l=1}^{L}\mathbb{Q}_{j-1,k} & \left(\sin\left(2\lambda_{k}t\right)\left(\mathbb{P}^{T}\mathbb{P}_{0}\mathbb{Q}_{0}^{T}\mathbb{Q}\right)_{k,l}\sin\left(2\lambda_{l}t\right)+\cos\left(2\lambda_{k}t\right)\left(\mathbb{P}^{T}\mathbb{P}_{0}\mathbb{Q}_{0}^{T}\mathbb{Q}\right)_{k,l}\cos\left(2\lambda_{l}t\right)\right)\mathbb{P}_{l,j}^{T},\\
\left\langle \left(c_{j}^{\dagger}-c_{j}^{\phantom{\dagger}}\right)\left(c_{j}^{\dagger}+c_{j}^{\phantom{\dagger}}\right)\right\rangle \nonumber \\
=-\sum_{k,l=1}^{L}\tilde{\mathbb{Q}}_{j,k} & \left(\sin\left(2\lambda_{k}t\right)\left(\mathbb{P}^{T}\mathbb{P}_{0}\mathbb{Q}_{0}^{T}\mathbb{Q}\right)_{k,l}\sin\left(2\lambda_{l}t\right)+\cos\left(2\lambda_{k}t\right)\left(\mathbb{P}^{T}\mathbb{P}_{0}\mathbb{Q}_{0}^{T}\mathbb{Q}\right)_{k,l}\cos\left(2\lambda_{l}t\right)\right)\mathbb{P}_{l,j}^{T}.
\end{align}

Finally, 
\begin{eqnarray}
{\cal E}_{j} & = & \frac{1}{2}J_{j-1}\sum_{k,l}\mathbb{Q}_{j-1,k}\left(\sin\left(2\lambda_{k}t\right)\left(\mathbb{P}^{T}\mathbb{P}_{0}\mathbb{Q}_{0}^{T}\mathbb{Q}\right)_{k,l}\sin\left(2\lambda_{l}t\right)+\cos\left(2\lambda_{k}t\right)\left(\mathbb{P}^{T}\mathbb{P}_{0}\mathbb{Q}_{0}^{T}\mathbb{Q}\right)_{k,l}\cos\left(2\lambda_{l}t\right)\right)\mathbb{P}_{l,j}^{T}\nonumber \\
 &  & -h_{j}\sum_{k,l}\mathbb{Q}_{j,k}\left(\sin\left(2\lambda_{k}t\right)\left(\mathbb{P}^{T}\mathbb{P}_{0}\mathbb{Q}_{0}^{T}\mathbb{Q}\right)_{k,l}\sin\left(2\lambda_{l}t\right)+\cos\left(2\lambda_{k}t\right)\left(\mathbb{P}^{T}\mathbb{P}_{0}\mathbb{Q}_{0}^{T}\mathbb{Q}\right)_{k,l}\cos\left(2\lambda_{l}t\right)\right)\mathbb{P}_{l,j}^{T}\nonumber \\
 &  & +\frac{1}{2}J_{j}\sum_{k,l}\mathbb{Q}_{j,k}\left(\sin\left(2\lambda_{k}t\right)\left(\mathbb{P}^{T}\mathbb{P}_{0}\mathbb{Q}_{0}^{T}\mathbb{Q}\right)_{k,l}\sin\left(2\lambda_{l}t\right)+\cos\left(2\lambda_{k}t\right)\left(\mathbb{P}^{T}\mathbb{P}_{0}\mathbb{Q}_{0}^{T}\mathbb{Q}\right)_{k,l}\cos\left(2\lambda_{l}t\right)\right)\mathbb{P}_{l,j+1}^{T},\label{eq:<energy>}\\
{\cal J}_{j} & = & J_{j-1}\left[h_{j}\sum_{k,l}\mathbb{Q}_{j-1,k}\left(\sin\left(2\lambda_{k}t\right)\left(\mathbb{P}^{T}\mathbb{P}_{0}\mathbb{Q}_{0}^{T}\mathbb{Q}\right)_{k,l}\cos\left(2\lambda_{l}t\right)-\cos\left(2\lambda_{k}t\right)\left(\mathbb{P}^{T}\mathbb{P}_{0}\mathbb{Q}_{0}^{T}\mathbb{Q}\right)_{k,l}\sin\left(2\lambda_{l}t\right)\right)\mathbb{Q}_{l,j}^{T}\right.\nonumber \\
 &  & \left.-h_{j-1}\sum_{k,l}\mathbb{P}_{j-1,k}\left(\cos\left(2\lambda_{k}t\right)\left(\mathbb{P}^{T}\mathbb{P}_{0}\mathbb{Q}_{0}^{T}\mathbb{Q}\right)_{k,l}\sin\left(2\lambda_{l}t\right)-\sin\left(2\lambda_{k}t\right)\left(\mathbb{P}^{T}\mathbb{P}_{0}\mathbb{Q}_{0}^{T}\mathbb{Q}\right)_{k,l}\cos\left(2\lambda_{l}t\right)\right)\mathbb{P}_{l,j}^{T}\right].\label{eq:<current>}
\end{eqnarray}

For completeness, the boundary terms are 
\begin{eqnarray}
{\cal E}_{1} & = & \frac{\left(-1\right)^{N+1}}{2}J_{L}\sum_{k,l}\mathbb{Q}_{L,k}\left(\sin\left(2\lambda_{k}t\right)\left(\mathbb{P}^{T}\mathbb{P}_{0}\mathbb{Q}_{0}^{T}\mathbb{Q}\right)_{k,l}\sin\left(2\lambda_{l}t\right)+\cos\left(2\lambda_{k}t\right)\left(\mathbb{P}^{T}\mathbb{P}_{0}\mathbb{Q}_{0}^{T}\mathbb{Q}\right)_{k,l}\cos\left(2\lambda_{l}t\right)\right)\mathbb{P}_{l,1}^{T}\nonumber \\
 &  & -h_{1}\sum_{k,l}\mathbb{Q}_{1,k}\left(\sin\left(2\lambda_{k}t\right)\left(\mathbb{P}^{T}\mathbb{P}_{0}\mathbb{Q}_{0}^{T}\mathbb{Q}\right)_{k,l}\sin\left(2\lambda_{l}t\right)+\cos\left(2\lambda_{k}t\right)\left(\mathbb{P}^{T}\mathbb{P}_{0}\mathbb{Q}_{0}^{T}\mathbb{Q}\right)_{k,l}\cos\left(2\lambda_{l}t\right)\right)\mathbb{P}_{l,1}^{T}\nonumber \\
 &  & +\frac{1}{2}J_{1}\sum_{k,l}\mathbb{Q}_{1,k}\left(\sin\left(2\lambda_{k}t\right)\left(\mathbb{P}^{T}\mathbb{P}_{0}\mathbb{Q}_{0}^{T}\mathbb{Q}\right)_{k,l}\sin\left(2\lambda_{l}t\right)+\cos\left(2\lambda_{k}t\right)\left(\mathbb{P}^{T}\mathbb{P}_{0}\mathbb{Q}_{0}^{T}\mathbb{Q}\right)_{k,l}\cos\left(2\lambda_{l}t\right)\right)\mathbb{P}_{l,2}^{T},\\
{\cal E}_{L} & = & \frac{1}{2}J_{L-1}\sum_{k,l}\mathbb{Q}_{L-1,k}\left(\sin\left(2\lambda_{k}t\right)\left(\mathbb{P}^{T}\mathbb{P}_{0}\mathbb{Q}_{0}^{T}\mathbb{Q}\right)_{k,l}\sin\left(2\lambda_{l}t\right)+\cos\left(2\lambda_{k}t\right)\left(\mathbb{P}^{T}\mathbb{P}_{0}\mathbb{Q}_{0}^{T}\mathbb{Q}\right)_{k,l}\cos\left(2\lambda_{l}t\right)\right)\mathbb{P}_{l,L}^{T}\nonumber \\
 &  & -h_{L}\sum_{k,l}\mathbb{Q}_{L,k}\left(\sin\left(2\lambda_{k}t\right)\left(\mathbb{P}^{T}\mathbb{P}_{0}\mathbb{Q}_{0}^{T}\mathbb{Q}\right)_{k,l}\sin\left(2\lambda_{l}t\right)+\cos\left(2\lambda_{k}t\right)\left(\mathbb{P}^{T}\mathbb{P}_{0}\mathbb{Q}_{0}^{T}\mathbb{Q}\right)_{k,l}\cos\left(2\lambda_{l}t\right)\right)\mathbb{P}_{l,L}^{T}\nonumber \\
 &  & +\frac{\left(-1\right)^{N+1}}{2}J_{L}\sum_{k,l}\mathbb{Q}_{L,k}\left(\sin\left(2\lambda_{k}t\right)\left(\mathbb{P}^{T}\mathbb{P}_{0}\mathbb{Q}_{0}^{T}\mathbb{Q}\right)_{k,l}\sin\left(2\lambda_{l}t\right)+\cos\left(2\lambda_{k}t\right)\left(\mathbb{P}^{T}\mathbb{P}_{0}\mathbb{Q}_{0}^{T}\mathbb{Q}\right)_{k,l}\cos\left(2\lambda_{l}t\right)\right)\mathbb{P}_{l,1}^{T},\\
{\cal J}_{1} & = & \left(-1\right)^{N+1}J_{L}\left[h_{1}\sum_{k,l}\mathbb{Q}_{L,k}\left(\sin\left(2\lambda_{k}t\right)\left(\mathbb{P}^{T}\mathbb{P}_{0}\mathbb{Q}_{0}^{T}\mathbb{Q}\right)_{k,l}\cos\left(2\lambda_{l}t\right)-\cos\left(2\lambda_{k}t\right)\left(\mathbb{P}^{T}\mathbb{P}_{0}\mathbb{Q}_{0}^{T}\mathbb{Q}\right)_{k,l}\sin\left(2\lambda_{l}t\right)\right)\mathbb{Q}_{l,1}^{T}\right.\nonumber \\
 &  & \left.-h_{L}\sum_{k,l}\mathbb{P}_{L,k}\left(\cos\left(2\lambda_{k}t\right)\left(\mathbb{P}^{T}\mathbb{P}_{0}\mathbb{Q}_{0}^{T}\mathbb{Q}\right)_{k,l}\sin\left(2\lambda_{l}t\right)-\sin\left(2\lambda_{k}t\right)\left(\mathbb{P}^{T}\mathbb{P}_{0}\mathbb{Q}_{0}^{T}\mathbb{Q}\right)_{k,l}\cos\left(2\lambda_{l}t\right)\right)\mathbb{P}_{l,1}^{T}\right].
\end{eqnarray}

In the main text, we are interested in the mean energy density above
the ground state of the post-quench Hamiltonian 
\begin{equation}
\delta{\cal E}_{j}(t)={\cal E}_{j}(t)-{\cal E}_{j}^{(\text{GS})},
\end{equation}
 where ${\cal E}_{j}^{(\text{GS})}=\left\langle \phi_{\text{GS}}\left|E_{j}\right|\phi_{\text{GS}}\right\rangle $
and $\left|\phi_{\text{GS}}\right\rangle $ is the ground state of
the post-quench Hamiltonian $H$. This is easily computed. Coming
back to \eqref{eq:E-density-i}, we then need the analogous of \eqref{eq:cxccyc}
which is $\left\langle \left(\mathbf{c}^{\dagger}+x\mathbf{c}\right)\left(\mathbf{c}^{\dagger T}+y\mathbf{c}^{T}\right)\right\rangle _{\text{GS}}=x\mathbb{P}_{x}\mathbb{P}_{y}^{T}.$
Then, 
\begin{eqnarray}
{\cal E}_{j}^{(\text{GS})} & = & \frac{1}{2}J_{j-1}\sum_{k}\mathbb{Q}_{j-1,k}\mathbb{P}_{j,k}-h_{j}\sum_{k}\mathbb{Q}_{j,k}\mathbb{P}_{j,k}+\frac{1}{2}J_{j}\sum_{k}\mathbb{Q}_{j,k}\mathbb{P}_{j+1,k}\nonumber \\
 & = & \frac{1}{2}J_{j-1}\left(\mathbb{Q}\mathbb{P}^{T}\right)_{j-1,j}-h_{j}\left(\mathbb{Q}\mathbb{P}^{T}\right)_{j,j}+\frac{1}{2}J_{j}\left(\mathbb{Q}\mathbb{P}^{T}\right)_{j,j+1},\\
{\cal E}_{1}^{(\text{GS})} & = & \frac{\left(-1\right)^{N+1}}{2}J_{L}\left(\mathbb{Q}\mathbb{P}^{T}\right)_{L,1}-h_{1}\left(\mathbb{Q}\mathbb{P}^{T}\right)_{1,1}+\frac{1}{2}J_{1}\left(\mathbb{Q}\mathbb{P}^{T}\right)_{1,2},\\
{\cal E}_{L}^{(\text{GS})} & = & \frac{1}{2}J_{L-1}\left(\mathbb{Q}\mathbb{P}^{T}\right)_{L-1,L}-h_{L}\left(\mathbb{Q}\mathbb{P}^{T}\right)_{L,L}+\frac{\left(-1\right)^{N+1}}{2}J_{L}\left(\mathbb{Q}\mathbb{P}^{T}\right)_{L,1}.
\end{eqnarray}

\section{The clean transverse-field Ising chain\label{sec:Z-clean}}

\subsection{Diagonalization}

The model Hamiltonian is 
\begin{equation}
H=-J\sum_{i=1}^{L}\sigma_{j}^{z}\sigma_{j+1}^{z}-h\sum_{j=1}^{L}\sigma_{j}^{x},\label{eq:H-Ising-clean}
\end{equation}
 where we are using periodic boundary conditions $\sigma_{L}^{x}\sigma_{L+1}^{x}=\sigma_{L}^{x}\sigma_{1}^{x}$.
From the mapping \eqref{eq:mapping}, then 
\begin{equation}
H=-J\sum_{j=1}^{L-1}\left(c_{j}^{\dagger}-c_{j}^{\phantom{\dagger}}\right)\left(c_{j+1}^{\dagger}+c_{j+1}^{\phantom{\dagger}}\right)+Je^{i\pi N}\left(c_{L}^{\dagger}-c_{L}^{\phantom{\dagger}}\right)\left(c_{1}^{\dagger}+c_{1}^{\phantom{\dagger}}\right)+h\sum_{j=1}^{L}\left(c_{j}^{\dagger}c_{j}^{\phantom{\dagger}}-c_{j}^{\phantom{\dagger}}c_{j}^{\dagger}\right).\label{eq:H-Ising-fermions}
\end{equation}
 Thus, the fermionic problem has periodic boundary conditions if the
total number of fermions is odd, and anti-periodic boundary conditions
otherwise.

Now, we use the Fourier transformation 
\begin{equation}
c_{j}=\sqrt{\frac{1}{L}}\sum_{n=1}^{L}e^{ik_{n}j}\gamma_{k_{n}}=\sqrt{\frac{1}{L}}\sum_{k}e^{ikj}\gamma_{k},\mbox{ where }k=k_{n}=\frac{\pi}{L}\left(2n-\frac{1+\left(-1\right)^{N+L}}{2}\right)-\pi,\ n=1,\dots,L.\label{eq:Fourier-Ising}
\end{equation}
 Then, 
\begin{eqnarray}
H & = & 2\sum_{0<k<\pi}\left(\begin{array}{cc}
\gamma_{k}^{\dagger} & \gamma_{-k}^{\phantom{\dagger}}\end{array}\right)\left(\begin{array}{cc}
h-J\cos k & -iJ\sin k\\
iJ\sin k & J\cos k-h
\end{array}\right)\left(\begin{array}{c}
\gamma_{k}^{\phantom{\dagger}}\\
\gamma_{-k}^{\dagger}
\end{array}\right)\nonumber \\
 &  & +\left(h+J\right)\delta_{\left(N+L+1\right)\text{mod}2,0}\left(\gamma_{\pi}^{\dagger}\gamma_{\pi}^{\phantom{\dagger}}-\gamma_{\pi}^{\phantom{\dagger}}\gamma_{\pi}^{\dagger}\right)+\left(h-J\right)\delta_{\left(N+1\right)\text{mod}2,0}\left(\gamma_{0}^{\dagger}\gamma_{0}^{\phantom{\dagger}}-\gamma_{0}^{\phantom{\dagger}}\gamma_{0}^{\dagger}\right).
\end{eqnarray}
 The modes $0$ and $\pi$ (when existing) are already diagonal. We
then diagonalize the remaining ones. This can be done by finding the
eigenvectors of the corresponding matrix (which is the Bogoliubov
transformation). Then, 
\begin{equation}
\mathbb{D}_{k}=\mathbb{V}_{k}^{*T}\left(\begin{array}{cc}
h-J\cos k & -iJ\sin k\\
iJ\sin k & J\cos k-h
\end{array}\right)\mathbb{V}_{k}=\left(\begin{array}{cc}
\epsilon_{k} & 0\\
0 & -\epsilon_{k}
\end{array}\right),
\end{equation}
 where $\mathbb{V}_{k}=\left(\begin{array}{cc}
\cos\theta & i\sin\theta\\
i\sin\theta & \cos\theta
\end{array}\right)$ is the eigenvector matrix and the dispersion relation is 
\begin{equation}
\omega_{k}=\sqrt{h^{2}-2hJ\cos k+J^{2}}.
\end{equation}
 The angle $\theta$ is such that 
\begin{equation}
\mathbb{D}_{k}=\left(\begin{array}{cc}
h\cos\left(2\theta\right)-J\cos\left(k+2\theta\right) & i\left(h\sin\left(2\theta\right)-J\sin\left(k+2\theta\right)\right)\\
-i\left(h\sin\left(2\theta\right)-J\sin\left(k+2\theta\right)\right) & J\cos\left(k+2\theta\right)-h\cos\left(2\theta\right)
\end{array}\right)=\left(\begin{array}{cc}
\omega_{k} & 0\\
0 & -\omega_{k}
\end{array}\right),
\end{equation}
 and thus, 
\begin{equation}
\tan\left(2\theta\right)=\frac{J\sin k}{h-J\cos k},\ \cos\left(2\theta\right)=\frac{h-J\cos k}{\omega_{k}},\ \sin\left(2\theta\right)=\frac{J\sin k}{\omega_{k}}.
\end{equation}
 Thus, the eigenfermions are 
\begin{equation}
\left(\begin{array}{c}
\eta_{k}^{\phantom{\dagger}}\\
\eta_{-k}^{\dagger}
\end{array}\right)=\mathbb{V}_{k}^{*T}\left(\begin{array}{c}
\gamma_{k}^{\phantom{\dagger}}\\
\gamma_{-k}^{\dagger}
\end{array}\right)=\left(\begin{array}{cc}
\cos\theta & -i\sin\theta\\
-i\sin\theta & \cos\theta
\end{array}\right)\left(\begin{array}{c}
\gamma_{k}^{\phantom{\dagger}}\\
\gamma_{-k}^{\dagger}
\end{array}\right)=\left(\begin{array}{c}
\cos\theta\gamma_{k}^{\phantom{\dagger}}-i\sin\theta\gamma_{-k}^{\dagger}\\
\cos\theta\gamma_{-k}^{\dagger}-i\sin\theta\gamma_{k}^{\phantom{\dagger}}
\end{array}\right).
\end{equation}
 The inverse transformation is 
\begin{equation}
\left(\begin{array}{c}
\gamma_{k}^{\phantom{\dagger}}\\
\gamma_{-k}^{\dagger}
\end{array}\right)=\mathbb{V}_{k}\left(\begin{array}{c}
\eta_{k}^{\phantom{\dagger}}\\
\eta_{-k}^{\dagger}
\end{array}\right)=\left(\begin{array}{c}
\cos\theta\eta_{k}^{\phantom{\dagger}}+i\sin\theta\eta_{-k}^{\dagger}\\
\cos\theta\eta_{-k}^{\dagger}+i\sin\theta\eta_{k}^{\phantom{\dagger}}
\end{array}\right).
\end{equation}
 Finally, the Hamiltonian is 
\begin{eqnarray}
H & = & \sum_{k}2\omega_{k}\left(\eta_{k}^{\dagger}\eta_{k}^{\phantom{\dagger}}-\frac{1}{2}\right).
\end{eqnarray}
 Notice that the modes $k=0$ and $\pi$ are trivial $\eta_{0}=\gamma_{0}$
and $\eta_{\pi}=\gamma_{\pi}$ and we are assuming, for simplicity,
that $h>J$. The ground-state energy is $E_{\text{GS}}=-\sum_{k}\omega_{k}$.

\subsection{The return probability amplitude}

We now want to compute the dynamical partition function 
\begin{equation}
Z\left(z\right)=\left\langle \text{GS}\left|e^{-iHz}\right|\text{GS}\right\rangle ,
\end{equation}
 where the post-quench Hamiltonian is $H=2\sum_{0<k<\pi}\omega_{k}\left(\eta_{k}^{\dagger}\eta_{k}^{\phantom{\dagger}}-\eta_{-k}^{\phantom{\dagger}}\eta_{-k}^{\dagger}\right)+H_{k=0}+H_{k=\pi}$,
with $H_{0,\pi}$ being the terms of the Hamiltonian concerning the
trivial modes ($\eta_{0,\pi}=\eta_{0,\pi}^{(0)}$) and, therefore,
contribute to $Z(z)$ with only a trivial dynamical phase. Notice
that the Fourier moments $k$ are not changed by the quench. Thus,
the problem simplifies in computing the $Z(z)$ for each $\pm k$
pair of modes: 
\begin{eqnarray}
Z\left(z\right) & = & \prod_{0<k<\pi}\left\langle \text{GS}_{-k}^{(0)},\text{GS}_{k}^{(0)}\left|e^{-i2\omega_{k}z\left(\eta_{k}^{\dagger}\eta_{k}^{\phantom{\dagger}}-\eta_{-k}^{\phantom{\dagger}}\eta_{-k}^{\dagger}\right)}\right|\text{GS}_{-k}^{(0)},\text{GS}_{k}^{(0)}\right\rangle \nonumber \\
 & = & \prod_{0<k<\pi}\left\langle \left[1+\left(x^{-1}-1\right)\eta_{k}^{\dagger}\eta_{k}^{\phantom{\dagger}}\right]\left[1+\left(x-1\right)\eta_{-k}^{\phantom{\dagger}}\eta_{-k}^{\dagger}\right]\right\rangle _{0},
\end{eqnarray}
 where $x=e^{2i\omega_{k}z}$. We need the relation between the new
and old eigen-fermions. 
\begin{equation}
\left(\begin{array}{c}
\eta_{k}^{\phantom{\dagger}}\\
\eta_{-k}^{\dagger}
\end{array}\right)=\mathbb{V}_{k}^{*T}\left(\begin{array}{c}
\gamma_{k}^{\phantom{\dagger}}\\
\gamma_{-k}^{\dagger}
\end{array}\right)=\mathbb{V}_{k}^{*T}\mathbb{V}_{k}^{(0)}\left(\begin{array}{c}
\eta_{k}^{(0)}\\
\eta_{-k}^{(0)\dagger}
\end{array}\right)=\left(\begin{array}{cc}
\cos\left(\Delta\theta\right) & -i\sin\left(\Delta\theta\right)\\
-i\sin\left(\Delta\theta\right) & \cos\left(\Delta\theta\right)
\end{array}\right)\left(\begin{array}{c}
\eta_{k}^{(0)}\\
\eta_{-k}^{(0)\dagger}
\end{array}\right),
\end{equation}
 where $\Delta\theta=\theta-\theta^{(0)}$. The ground-state mean
value of 
\begin{equation}
1+\left(x^{-1}-1\right)\eta_{k}^{\dagger}\eta_{k}^{\phantom{\dagger}}+\left(x-1\right)\eta_{-k}^{\phantom{\dagger}}\eta_{-k}^{\dagger}+\left(x^{-1}-1\right)\left(x-1\right)\eta_{k}^{\dagger}\eta_{k}^{\phantom{\dagger}}\eta_{-k}^{\phantom{\dagger}}\eta_{-k}^{\dagger},
\end{equation}
 is 
\begin{equation}
1+\left(x^{-1}-1\right)\sin^{2}\Delta\theta+\left(x-1\right)\cos^{2}\Delta\theta=x^{-1}\sin^{2}\Delta\theta+x\cos^{2}\Delta\theta.
\end{equation}
 Inserting the trivial dynamical phases from the trivial $0$ and
$\pi$ modes, then 
\begin{eqnarray}
Z(z) & = & e^{-iE_{\text{GS}}z}\prod_{0<k<\pi}\left(1-\frac{1}{2}\left(1-e^{-4i\omega_{k}z}\right)\left(1-\frac{hh_{0}-\left(Jh_{0}+hJ_{0}\right)\cos k+JJ_{0}}{\omega_{k}\omega_{k}^{(0)}}\right)\right).
\end{eqnarray}
 The zeros of $Z(z)$, $z^{*}$, are given by 
\begin{equation}
2\omega_{k}z^{*}=\left(m+\frac{1}{2}\right)\pi-\frac{i}{2}\ln\tan^{2}\Delta\theta,
\end{equation}
 with $m\in\mathbb{N}$. The zeros pierce the real-time axis if $\tan^{2}\Delta\theta$
crosses the value $1$ for $0<k<\pi$. This is only possible if 
\begin{equation}
\left(h-J\right)\left(h_{0}-J_{0}\right)<0.
\end{equation}
 In other words, the zeros of $Z\left(z\right)$ crosses the real-time
axis only if the quench crosses the equilibrium quantum phase transition.

Further manipulations allows us to rewrite the zeros as 
\begin{equation}
2\omega_{k}z^{*}=\left(m+\frac{1}{2}\right)\pi+\frac{i}{2}\ln\left(\frac{\omega_{k}\omega_{k}^{(0)}+\left(hh_{0}-\left(Jh_{0}+hJ_{0}\right)\cos k+JJ_{0}\right)}{\omega_{k}\omega_{k}^{(0)}-\left(hh_{0}-\left(Jh_{0}+hJ_{0}\right)\cos k+JJ_{0}\right)}\right),\label{eq:zeros-clean-Ising}
\end{equation}
 which recovers Eq.~\eqref{eq:YLF-clean} of the main text. How many
zeros are there in a single accumulation line? From \eqref{eq:zeros-clean-Ising},
it is just the total number of $k$'s between $0$ and $\pi$ (excluding
$0$ and $\pi$). From \eqref{eq:Fourier-Ising}, it is simply the
largest integer less than $\frac{L+(1+\left(-1\right)^{N})/2}{2}$
(recall $k=\pi$ is excluded). Thus, 
\begin{equation}
n_{\text{zeros}}=\frac{1}{2}\left(L-1+\left(-1\right)^{N}\left(1-L\text{mod}2\right)\right)=\frac{1}{2}\left(L-1+\left(-1\right)^{N}\left(\frac{1+\left(-1\right)^{L}}{2}\right)\right).
\end{equation}

\begin{figure}[h]
\begin{centering}
\includegraphics[clip,width=6cm]{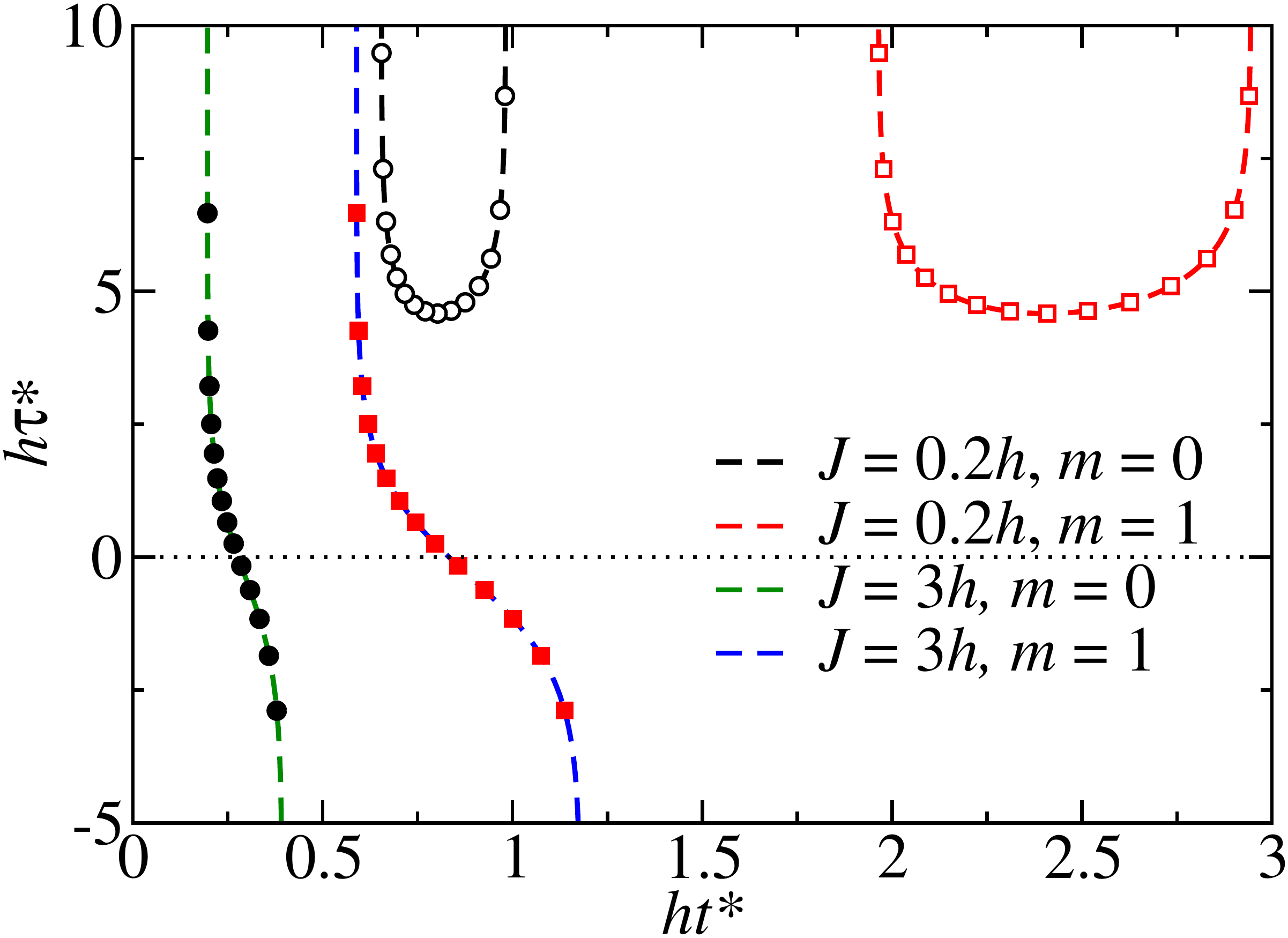}
\par\end{centering}
\caption{\label{fig:clean-zeros} Zeros of the dynamical partition function
for a clean chain of $30$ sites long with periodic boundary conditions
(symbols) and the corresponding accumulation lines in the thermodynamic
limit (dashed lines). The zeros are those given by Eq.~\eqref{eq:YLF-clean}
of the main text {[}equivalent to Eq.~\eqref{eq:zeros-clean-Ising}{]}.
The lines are simply the zeros of the same equation with $L\rightarrow\infty$.
All quantum quenches are from $h_{0}=\infty$ to finite $h$. We also
show two accumulation lines $m=0$ (black and dark green) and $m=1$
(red and blue). In one chain, the coupling constants are equal to
$J=0.2h$ (accumulation lines in the upper imaginary plane, black
and red; open symbols), and equal to $J=3h$ in the other (accumulation
lines piercing the real-time axis, dark green and blue; closed symbols).}
\end{figure}

As discussed in the main text, the zeros of $Z$ accumulate in lines
which can only pierce the real-time axis if the equilibrium quantum
phase transition is crossed by the quench. This is illustrated in
Fig.~\ref{fig:clean-zeros}. In the thermodynamic limit, the imaginary
part of $z^{*}$ vanishes when $hh_{0}-\left(Jh_{0}+hJ_{0}\right)\cos k+JJ_{0}=0$.
Thus, the associated momentum $q$ is given by 
\begin{equation}
\cos q=\frac{hh_{0}+JJ_{0}}{Jh_{0}+hJ_{0}}.
\end{equation}
 The corresponding zero is a real number and equals 
\begin{equation}
t^{*}=\frac{\left(2n+1\right)\pi}{4\omega_{q}}=\frac{\left(2n+1\right)\pi}{4}\sqrt{\frac{hJ_{0}+Jh_{0}}{\left(h^{2}-J^{2}\right)\left(hJ_{0}-Jh_{0}\right)}}.
\end{equation}

For finite $L$, none of the momenta $k_{n}$ in \eqref{eq:Fourier-Ising}
matches $q$ in general. However, in the worst case, the closest $k_{n}$
to $q$ is far by $\delta k=\frac{2\pi}{L}$. Thus, in the large-$L$
regime, the imaginary part of the closest zero to the real-time axis
vanishes as $\tau^{*}\propto\delta k\propto L^{-1}$. Although we
have explicitly derived this result for a chain with periodic boundary
conditions, we expect it to be valid for chains with open boundary
conditions, as well. A detailed analysis will be reported elsewhere.

\section{A single Rare Region}

\begin{figure}
\begin{centering}
\includegraphics[clip,width=6cm]{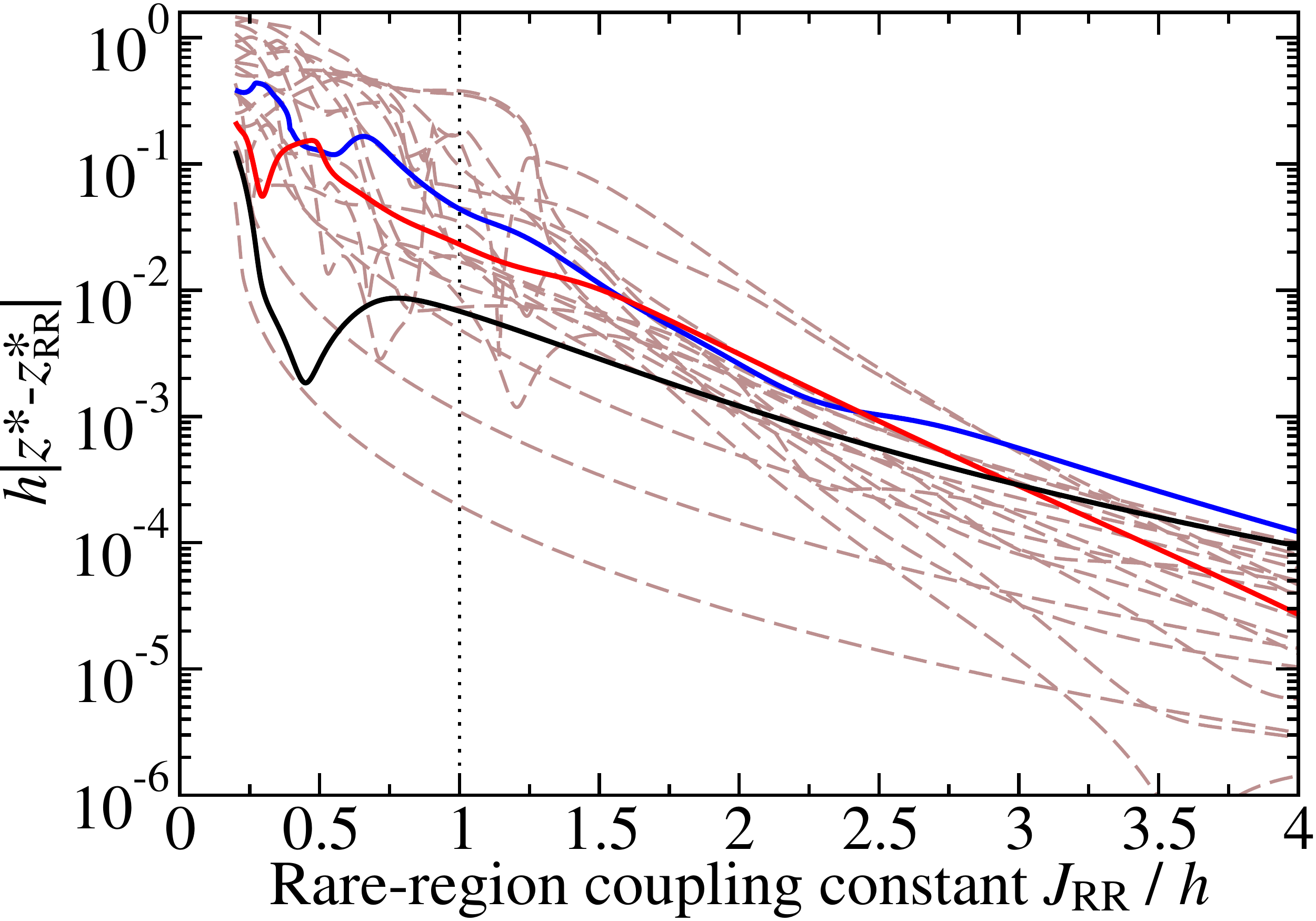}
\par\end{centering}
\caption{\label{fig:comparison}The absolute difference between the rare-region-induced
set of zeros near the real-time axis $z^{*}$ and those of a decoupled
rare region $z_{\text{RR}}^{*}$ shown in Fig.~\hyperref[fig:YLF]{\ref{fig:YLF}(b)}
of the main text. In total, we have compared $21$ zeros. The difference
is plotted as a function of the rare-region coupling constant $J_{\text{RR}}$.
The three solid lines correspond to the $3$ closest zeros to the
real axis, time instants $ht\approx0.6$ (black), $1.7$ (red), and
$2.8$ (blue).}
\end{figure}

In the main text, we showed that the Yang-Lee-Fisher (YLF) zeros of
the dynamical partition function $Z(z)$, Eq.~\eqref{eq:Z} of the
main text, perform nontrivial paths in the complex-time plane when
a Rare Region (RR) appears in the system. One set of the zeros remains
in the upper complex plane and the other migrates close to the real
time axis. In addition, we showed that this second set of zeros $\left\{ z^{*}\right\} $
and the singular part of the dynamical free-energy $f(t)$ are well
described by those same quantities of a decoupled RR undergoing the
same quantum quench, $\left\{ z_{\text{RR}}^{*}\right\} $ and $f_{\text{RR}}$,
respectively. In this section, we quantify this result. We compute
the difference between $z^{*}$ and $z_{\text{RR}}^{*}$ as a function
of $J_{\text{RR}}$ (the coupling constant inside the RR). This is
shown in Fig.~\ref{fig:comparison}. As can be seen, the difference
vanishes exponentially (with possible algebraic and nontrivial oscillatory
corrections) with $J_{\text{RR}}$. Evidently, this behavior becomes
evident when $J_{\text{RR}}$ becomes greater than $h$, as expected.

\section{Two Rare Regions}

In the main text, we have shown that the quantum-quench-induced excitations
are localized in each RR if they are sufficiently far apart from each
other (see Fig.~\ref{fig:E-J-2RR} of the main text). In this section,
we give further evidence of this result. 

\begin{figure}[t]
\begin{centering}
\includegraphics[clip,width=6cm]{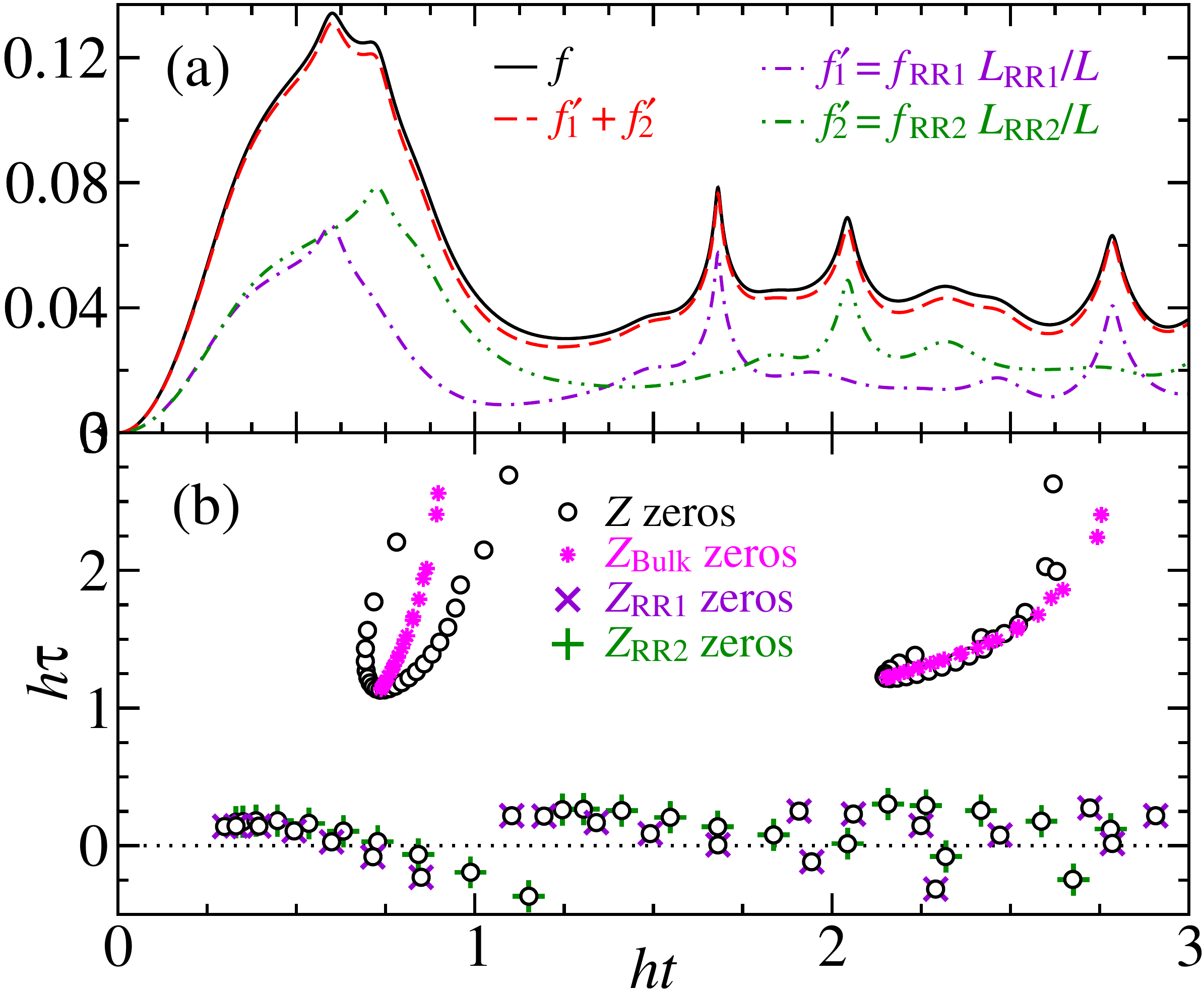}
\par\end{centering}
\caption{\label{fig:YLF-2RR}(a) The dynamical free energy $f$ as a function
of the real time $t$ and (b) the associated Yang-Lee-Fisher zeros
of the return probability amplitude $Z(z)$ with $z=t+i\tau$. The
quantum quench of the Hamiltonian \eqref{eq:H} (of the main text)
is from $h_{0}=\infty$ to $h$. The chain is $70$ sites long with
periodic boundary conditions. The bulk coupling constant is $J_{\text{B}}=0.2h$.
The chain has two rare regions. The first (second) one comprises sites
$1$ to $8$ (40 to 50). Thus, $L_{\text{RR1}}=8$ ($L_{\text{RR2}}=11$).
The corresponding coupling constant is $J_{\text{RR}1}=3h$ ($J_{\text{RR}2}=2.5h$).
The free energy and the corresponding zeros of the decoupled rare
regions are plotted as well (see text).}
\end{figure}

In Fig.~\ref{fig:YLF-2RR}, we plot the dynamical free energy $f(t)$
{[}panel (a), continuous black line{]} and the corresponding zeros
of $Z(z)$ {[}panel (b), open black circles{]} for a chain of $L=70$
sites long with periodic boundary conditions. The quench is from $h_{0}=\infty$
to a finite $h$. The bulk coupling constant is $J_{\text{B}}=0.2h$.
The chains has two RRs. The first one is $L_{\text{RR}1}=8$ sites
long and its coupling is $J_{\text{RR}1}=3h$, and the second one
is $L_{\text{RR}2}=11$ sites long and its coupling is $J_{\text{RR}2}=2.5h$.

As for a single RR (see Fig.~\ref{fig:YLF} of the main text), the
zeros group themselves in two sets: one up in the positive complex
plane and the other near the real-time axis. This second set of zeros
is well approximated by decoupled RRs. We plot the dynamical free
energy $f_{\text{RR}1}$ and the associated YLF zeros of the first
decoupled RR as a purple dash-dotted line and purple $\times$ symbols
in panel (a) and (b) of Fig.~\ref{fig:YLF-2RR}, respectively. Likewise
for the second RR.

Interestingly, the zeros of the decoupled RRs reproduce accurately
the set of zeros which accumulate in lines piercing the real-time
axis. In addition, the superposition (simple sum) of the free energies
of the decoupled RRs (appropriately reweighted by $L_{\text{RR}}/L$)
accurately reproduce the singular part of the free energy $f$ {[}see
red dashed line of \hyperref[fig:YLF-2RR]{\ref{fig:YLF-2RR}(a)}{]}.

On the other hand, the set of zeros in the upper complex-time plane,
which are, presumably, due to the bulk, is not well approximate by
a decoupled bulk. The magenta stars in Fig.~\hyperref[fig:YLF-2RR]{\ref{fig:YLF-2RR}(b)}
are the YLF zeros of the decoupled bulk, i.e., the zeros corresponding
to two open boundary chains of sizes $31$ and $20$ undergoing the
same quantum quench from $h_{0}\rightarrow\infty$ to $h$ where the
coupling constant of these chains is $J_{\text{B}}=0.2h$. This means
that that analytic part of $f$ cannot be well described by decoupled
bulk and rare regions.

\section{The case of extreme quenches\label{sec:extreme-quench}}

Consider the simple quantum quench from 
\begin{equation}
H_{0}=-h_{0}\sum_{i=1}^{L}\sigma_{i}^{x}\mbox{ to }H=-\sum_{i=1}^{L}J_{i}\sigma_{i}^{z}\sigma_{i+1}^{z}.
\end{equation}
The initial state is $\left|\psi_{0}\right\rangle =2^{-\frac{L}{2}}\left|\left\{ s_{i}\right\} \right\rangle $,
with $s_{i}=\pm1$ and the set $\left\{ s_{i}\right\} $ covers all
the $2^{L}$ possible spin configurations. Thus, 
\begin{equation}
Z=2^{-L}\sum_{\left\{ s_{i}\right\} }\sum_{\left\{ t_{i}\right\} }\left\langle \left\{ t_{i}\right\} \left|e^{-iHz}\right|\left\{ s_{i}\right\} \right\rangle =2^{-L}\sum_{\left\{ s_{i}\right\} }e^{-i\left(-\sum_{k}J_{k}s_{k}s_{k+1}\right)z},
\end{equation}
 which is the partition function of the classical Ising chain in zero
longitudinal field. This can be computed via transfer matrix:

\begin{eqnarray}
Z & = & 2^{-L}\text{Tr}\left[\left(\begin{array}{cc}
e^{iJ_{1}z} & e^{-iJ_{1}z}\\
e^{-iJ_{1}z} & e^{iJ_{1}z}
\end{array}\right)\left(\begin{array}{cc}
e^{iJ_{2}z} & e^{-iJ_{2}z}\\
e^{-iJ_{2}z} & e^{iJ_{2}z}
\end{array}\right)\dots\left(\begin{array}{cc}
e^{iJ_{L}z} & e^{-iJ_{L}z}\\
e^{-iJ_{L}z} & e^{iJ_{L}z}
\end{array}\right)\right]\nonumber \\
 & = & \prod_{k=1}^{L}\cos\left(J_{k}z\right)+i^{L}\prod_{k=1}^{L}\sin\left(J_{k}z\right).\label{eq:Z-extreme}
\end{eqnarray}
 When at least one coupling is vanishing (as for open boundary condition
or as for the percolation problem), the imaginary part vanishes identically
and \eqref{eq:Z-extreme} recovers Eq.~\eqref{eq:fsing-extreme}
of the main text.
\end{document}